\documentclass[twocolumn]{aastex701}

\newcommand{\HA}{H$\alpha$}					

\shorttitle{SHIELD {\HA} SFRs}
\shortauthors{Shepley et al.}

\begin{document}

\title{H$\alpha$ Star Formation Rates of the SHIELD Dwarf Galaxies }

\correspondingauthor{John Salzer}
\email{josalzer@iu.edu}

\author{Madeline Shepley}
\affiliation{Department of Astronomy, Indiana University, 727 East Third Street, Bloomington, IN 47405, USA}
\affiliation{Department of Chemistry and Geoscience, Olivet Nazarene University, 1 University Avenue, Bourbonnais, IL  60914, USA}
\email{shepley2698@gmail.com}

\author[0009-0001-1540-3246]{David G. Gormanous}
\affiliation{Department of Astronomy, Indiana University, 727 East Third Street, Bloomington, IN 47405, USA}
\email{dgormano@iu.edu}

\author[0000-0001-8483-603X]{John J. Salzer}
\affiliation{Department of Astronomy, Indiana University, 727 East Third Street, Bloomington, IN 47405, USA}
\email{josalzer@iu.edu}

\author[0000-0001-5247-1371]{Nathalie Haurberg}
\affiliation{Department of Physics and Astronomy, Knox College, 2 East South Street, Galesburg, IL 61401, USA}
\email{nhaurber@knox.edu}

\author[0000-0001-9165-8905]{Steven Janowiecki}
\affiliation{Hobby-Eberly Telescope, McDonald Observatory, University of Texas, Austin, TX 78712, USA}
\email{steven.janowiecki@gmail.com}

\author[0000-0002-1821-7019]{John M. Cannon}
\affiliation{Department of Physics and Astronomy, Macalester College, Saint Paul, MN 55105, USA}
\email{jcannon@macalester.edu}

\author[0000-0002-9798-5111]{Elizabeth A. K. Adams}
\affiliation{ASTRON, The Netherlands Institute for Radio Astronomy, Oude Hoogeveensedijk 4, 7991 PD, Dwingeloo, The Netherlands}
\affiliation{Kapteyn Astronomical Institute, University of Groningen, Postbus 800, 9700 AV Groningen, The Netherlands} 
\email{adams@astron.nl}

\author[0000-0001-5334-5166]{Martha P. Haynes}
\affiliation{Center for Astrophysics and Planetary Science, Space Sciences Building, Cornell University, Ithaca, NY 14853, USA}
\email{haynes@astro.cornell.edu}

\author[0000-0002-2954-8622]{Alec S. Hirschauer}
\affiliation{Department of Physics \& Engineering Physics, Morgan State University, 1700 East Cold Spring Lane, Baltimore, MD 21251, USA}
\email{alechirschauer@gmail.com}

\author[0000-0001-5538-2614]{Kristen B. W. McQuinn}
\affiliation{Space Telescope Science Institute, 3700 San Martin Drive, Baltimore, MD 21218, USA}
\email{kmcquinn@stsci.edu}

\author[0000-0001-8283-4591]{Katherine L. Rhode}
\affiliation{Department of Astronomy, Indiana University, 727 East Third Street, Bloomington, IN 47405, USA}
\email{krhode@iu.edu}

\author[0000-0003-0605-8732]{Evan D. Skillman}
\affiliation{Minnesota Institute for Astrophysics, School of Physics and Astronomy, University of Minnesota, 116 Church Street, S.E., Minneapolis, MN 55455, USA}
\email{skill001@umn.edu}

\begin{abstract}
The Survey of \ion{H}{1} in Extremely Low-mass Dwarfs (SHIELD) galaxy sample is an \ion{H}{1}-selected catalog of galaxies that occupy the extreme low-mass end of the \ion{H}{1} mass function.  Our team is carrying out a comprehensive multi-wavelength observational program targeting these gas-rich, star-forming dwarf galaxies. The current paper presents H$\alpha$ narrowband observations of the full SHIELD sample. We detect 55 of the 82 galaxies (67\%) with nebular H$\alpha$ emission, and infer sensitive upper limits to the H$\alpha$ fluxes for the 27 non-detections.  We derive H$\alpha$ luminosities and star formation rates (SFRs), or upper limits, for all 82 SHIELD galaxies.  We also present, for the first time, deep broadband and continuum-subtracted H$\alpha$ images plus a summary of the properties for the entire sample.  Our analysis of the H$\alpha$ data re-affirms previous results which show that at low levels of star formation (log(SFR$_{H\alpha}$) $<$ $-$2.5), H$\alpha$ becomes an increasingly poor tracer of the recent average SFR as measured by far-UV (FUV) luminosities.  However, we show that this effect is not strictly correlated with the mass of the galaxy.   Some very low-mass SHIELD galaxies with higher H$\alpha$-based SFRs show good agreement with SFR$_{FUV}$.  The SHIELD galaxies show no correlation between H$\alpha$ SFR and gas mass fraction, but reveal a modest correlation between H$\alpha$ specific SFR (sSFR$_{H\alpha}$) and gas mass fraction.  Overall, the star formation efficiency exhibited by the SHIELD dwarfs is rather low, despite the gas-rich nature of the sample (68\% have M$_{gas}$ $>$ M$_{stellar}$).  

\end{abstract}



\section{Introduction} \label{sec:intro}

Dwarf galaxies represent important laboratories for studying the physical processes that regulate galaxy evolution. Their shallow gravitational potentials, relatively simple structures, and low masses make them particularly sensitive to the effects of star formation and feedback. Because they constitute the most numerous class of galaxies in the local universe, understanding their properties is essential for developing a complete picture of galaxy evolution across the full mass spectrum. In particular, studies of dwarf galaxies offer valuable insight into how gas is converted into stars under conditions that differ substantially from those found in more massive spiral systems.

Measuring the rate of star formation in low-mass galaxies remains one of the challenges in the study of dwarf galaxy evolution. Different star-formation tracers probe different stellar populations and timescales, and these tracers can yield significantly different results in galaxies with very low levels of star-formation activity \citep[e.g.,][]{Lee_2009b, Walter_2008, hunter2012, Roychowdhury_2014, teich2016, ott2012, gormanous26}. Far-ultraviolet (FUV) emission traces the integrated light from young stellar populations over timescales of roughly 100 Myr, whereas H$\alpha$ emission originates from ionized gas surrounding the most massive O-type stars and probes star formation on timescales of only $\lesssim$10 Myr. Numerous studies have shown that H$\alpha$-derived star-formation rates (SFRs) for low-mass dwarf galaxies systematically underestimate actual SFRs derived using FUV fluxes, suggesting that the assumptions underlying standard SFR calibrations may break down in this regime \citep{Lee_2009b, Hunter_2010, fumagalli2011, weisz2012, Roychowdhury_2014, teich2016, gormanous26}.

Several explanations have been proposed for the divergence between H$\alpha$ and FUV SFRs in low-mass systems, including stochastic sampling of the upper end of the stellar initial mass function \citep{fumagalli2011, dasilva2012}, bursty star-formation histories \citep{weisz2012}, the leakage of ionizing photons from star-forming regions \citep{oey2007} and a non-constant initial mass function \citep[IMF,][]{meurer2009, Gunawardhana2011}. Distinguishing among these possibilities and determining the relevant physical scales at which these effects become important remain key goals. Consequently, obtaining high-quality H$\alpha$ observations for statistically meaningful samples of extremely low-mass galaxies is essential for understanding how current star formation proceeds in environments where stochastic effects may dominate.

The Survey of \ion{H}{1} in Extremely Low-mass Dwarfs \citep[SHIELD,][]{cannon2011} provides a 
sample of dwarf galaxies that is uniquely positioned to address these questions, targeting the low-mass, gas-rich regime where these effects begin to matter.  SHIELD is an \ion{H}{1}-selected galaxy sample that probes the extreme low gas mass end of the local galaxy population.  A detailed description of the sample selection, completeness, and ancillary data products is given in \citet{gormanous26} and summarized below. The SHIELD galaxies have been studied across a broad range of wavelengths in order to characterize their gas content, stellar populations, star-formation properties, and chemical evolution.

The focus of the current paper is to present the results of our comprehensive narrowband imaging program and to explore the H$\alpha$ SFRs of the full SHIELD sample. This work expands upon an earlier companion paper that studied the FUV SFRs for the full sample of 82 SHIELD galaxies \citep{gormanous26}.

This paper is laid out as follows: Section~\ref{sec:background} introduces the SHIELD galaxy sample, gives a summary of previous SHIELD-based studies, and describes our H$\alpha$ observations. Section~\ref{sec:sfrdet} presents the methodology for our H$\alpha$ flux measurements and SFR calculations.  Section~\ref{sec:results} gives the results of our narrowband imaging, discusses the global properties of the SHIELD galaxies, exhibits broadband and H$\alpha$ images for the full SHIELD sample, and carries out an analysis of H$\alpha$ SFRs, highlighting the properties of both the SHIELD galaxies and an extensive comparison sample. Section~\ref{sec:summary} summarizes the main results of this paper.

\section{Sample Description and Observation} \label{sec:background}

\subsection{The SHIELD Sample of Dwarf Galaxies}\label{shield}

A fundamental goal of the SHIELD project has been to broadly refine our knowledge of the properties of gas-bearing dwarf galaxies in the local universe (5-12 Mpc away). The full SHIELD catalog includes 82 galaxies, which represent a volumetrically complete sample of the sky within the given selection constraints listed below. The SHIELD galaxies are an \ion{H}{1}-selected sample, chosen from the Arecibo Legacy Fast ALFA (ALFALFA) survey \citep{giovanelli2005, haynes2011, haynes2018}. We note that the ALFALFA-detected galaxies populate the low-mass end (e.g., M$_{HI}$ $<$ 10$^7$ M$_\odot$) of the \ion{H}{1} mass function at a significantly improved level compared to previous \ion{H}{1} surveys \citep{martin2010, jones2018}, making this survey essential for our understanding of the properties of the lowest-mass gas-bearing dwarf galaxies. The primary selection criteria of the SHIELD sample include: (i) \ion{H}{1} gas mass below 1.6 $\times$ 10$^7$ M$_\odot$, (ii) \ion{H}{1} velocity widths (FWHM) of below 65 km s$^{-1}$ (to exclude massive but gas-poor galaxies),  (iii) the presence of a detected stellar population as indicated by stellar emission visible in existing wide-field imaging surveys such as SDSS, and (iv) distances less than $\sim$11 Mpc to ensure that the galaxies were close enough for detailed analysis \citep{cannon2011, mcquinn2021}.  In general, nearby galaxies (distances less than $\sim$4 Mpc) that satisfied these selection criteria but were already included in existing \ion{H}{1} surveys were not included in SHIELD.

A central goal of the SHIELD project is to determine how gas dynamics, star formation, and feedback operate in extremely low-mass, gas-rich galaxies, and how these conditions differ from those in more massive systems, thereby constraining models of galaxy formation and evolution. To this end, SHIELD galaxies have been studied across a broad wavelength range to quantify their gas content, stellar populations, star formation properties, and chemical enrichment \citep{cannon2011, haurberg2015, mcquinn2015a, teich2016, mcnichols2016, mcquinn2021, mcquinn2022, gormanous26}. Existing data include ground-based optical broadband and narrowband H$\alpha$ imaging; optical spectroscopy of embedded \ion{H}{2} regions; multi-configuration VLA and WSRT \ion{H}{1} synthesis imaging; Spitzer/IRAC 3.6~$\mu$m imaging; and GALEX NUV and FUV measurements. A companion paper presents FUV-based star formation rates for the full SHIELD sample \citep{gormanous26}. Approximately half of the SHIELD galaxies possess HST imaging enabling robust tip of the red giant branch (TRGB) distance measurements \citep{mcquinn2014,mcquinn2021}.

\subsection{Narrowband H$\alpha$ Imaging Observations}\label{observation}
A long-standing goal of the SHIELD collaboration has been to obtain narrowband H$\alpha$ images for the entire sample of 82 galaxies.  The scientific motivations for carrying out such an observing program are many, and include (i) measuring the current star-formation rates (SFRs) for all of the SHIELD galaxies; (ii) using the H$\alpha$ images to map out the best locations within the SHIELD galaxies for carrying out spectroscopic observations with the goal of measuring nebular abundances; and (iii) allowing for a direct comparison between the deep, high-resolution \ion{H}{1} maps \citep[e.g.,][]{teich2016,mcnichols2016} and the H$\alpha$ images in order to assess the degree to which the surface density of the cold gas tracks the locations of current star formation.

H$\alpha$ images of the SHIELD galaxies were obtained over the course of several years. The initial 12 SHIELD galaxies were observed using the WIYN Observatory\footnote[1]{The WIYN Observatory is a joint facility of the University of Wisconsin-Madison, Indiana University, and NSF/NOIRLab.} 3.5-meter telescope. The observations, image analysis and measurements are described in detail in \citet{haurberg2015}.

The remaining 70 galaxies were observed using the WIYN 0.9-meter telescope and the Half-Degree Imager (HDI) camera.  These data were obtained over four observing runs as follows: (i) 31 galaxies were observed over 5 nights in February 2014 by Steven Janowiecki and Alec Hirschauer; (ii) 28 galaxies were observed over 4 nights in March 2016 by Nathalie Haurberg; (iii) nine galaxies were observed over 3 nights in September 2017 by John Salzer; (iv) 13 galaxies were observed over 3 nights in April 2018 by Nathalie Haurberg.  We note that the total number of objects observed during these latter four runs total more than 70 galaxies because 11 galaxies for which no H$\alpha$ emission was detected in earlier observing runs were re-observed in April 2018.  All of these latter galaxies remained undetected in the follow-up observations.

All 70 SHIELD galaxies observed with the WIYN 0.9-meter telescope were imaged through two fiters: a broadband Johnson-Cousins R filter and a H$\alpha$ narrowband filter centered on $\lambda$ = 6577 \AA\ with a FWHM of $\Delta\lambda$ = 59 \AA. For each galaxy, one R-band image was taken with an exposure time of 300 seconds, and two H$\alpha$ images were obtained that bracketed the R-band image and had exposure times of 1200 seconds each. To calibrate the science frames, bias images and dome flats were taken each day. Dark images were obtained periodically, but since the dark level of the HDI CCD was always found to be negligible, these darks were not used during the image processing.

On every night that data were obtained we also observed several broadband photometric standard stars \citep[e.g.,][]{landolt1992} to provide calibration for the R-band magnitudes, as well as spectrophotometric standard stars \citep[e.g.,][]{massey1988} for calibrating the H$\alpha$ fluxes.

Our narrowband images are quite sensitive to the appearance of \ion{H}{2} regions ionized by massive stars - which serves both goals of measuring H$\alpha$ SFRs and providing targets for nebular spectroscopy.  However, since most of our images were obtained with a small telescope (aperture of 0.9 m), our images are not particularly sensitive to diffuse/low surface brightness emission.  Hence, the absence of detected H$\alpha$ does not imply that low surface brightness H$\alpha$ emission does not exist.  As an example, compare the similar H$\alpha$ imaging of Leo~P that led to the discovery of its \ion{H}{2} region \citep{rhode2013} with the deeper MUSE/IFU imaging obtained by \citet{evans2019} that revealed extansive low surface brightness H$\alpha$ bubbles. 


\section{Data Processing and Measurements}\label{sec:sfrdet}

\subsection{Image Processing}\label{processing}
The SHIELD images were processed by applying standard reduction strategies.  All images had the bias level removed using the overscan region, after which the mean bias image was subtracted to account for any 2D structure in the bias.  Average-combined dome flats for each night were created, normalized, and used to correct the science images for pixel-to-pixel sensitivity variations.  An illumination correction image was created by performing a median combine on all of the science images obtained through both the R-band and NB filters during a given observing run.  The application of the normalized illumination correction images to the science images resulted in flat backgrounds across the full extent of the final processed images.

After the removal of all detector-dependent signatures from the data, the processed science images for each target galaxy were run through a series of scripts that performed the following tasks: (i) alignment of the three images for each galaxy to a common center, (ii) smoothing the images so that all three had the same stellar point-spread function width (FWHM), (iii) using bright stars to scale the flux level in the R-band image to match that in the combined narrowband H$\alpha$ image, and (iv) subtracting the scaled R-band image from the H$\alpha$ image, yielding a continuum-subtracted image of the SHIELD galaxy.  Our continuum-subtracted H$\alpha$ images are shown in Section~\ref{sec:images}.

\begin{deluxetable*}{cccccc}
\digitalasset
\tabletypesize{\scriptsize}
\tablecaption{Measured H$\alpha$ Fluxes for SHIELD Galaxies\label{tab:haflux}}
\tablehead{
  \colhead{AGC \#} &   \colhead{RA} &   \colhead{Dec} &   \colhead{m$_R$} &   \colhead{H$\alpha$ Flux} &   \colhead{A$_{H\alpha}$}   \\
& (J2000) & (J2000) & mag & 10$^{-14}$ erg s$^{-1}$ cm$^{-2}$  & mag \\
(1) & (2) & (3) & (4) & (5) & (6) 
} 
\startdata 
 102728 &   0.08917 & 31.02194 & 18.97 $\pm$ 0.08 &  (0.030 $\pm$ 0.010) & 0.10 \\
 103722 &   3.69167 & 10.81306 & 17.25 $\pm$ 0.05 &  0.445 $\pm$ 0.031 & 0.22 \\
 104208 &  10.42708 & 12.99222 & 19.55 $\pm$ 0.17 &  (0.030 $\pm$ 0.010) & 0.23 \\
 110482 &  25.57208 & 26.36667 & 15.65 $\pm$ 0.03 &  1.876 $\pm$ 0.039 & 0.21 \\
 111164 &  30.04208 & 28.83111 & 16.61 $\pm$ 0.03 &  1.177 $\pm$ 0.017 & 0.13 \\
 111946 &  26.67583 & 26.80139 & 17.26 $\pm$ 0.03 &  0.437 $\pm$ 0.020 & 0.19 \\
 111977 &  28.83417 & 27.95389 & 15.43 $\pm$ 0.03 &  1.360 $\pm$ 0.029 & 0.16 \\
 112503 &  24.50125 & 14.98278 & 16.73 $\pm$ 0.03 &  2.272 $\pm$ 0.073 & 0.13 \\
 112505 &  25.04000 & 15.94000 & 20.68 $\pm$ 0.34 &  (0.030 $\pm$ 0.010) & 0.15 \\
 112521 &  25.28333 & 27.32222 & 17.62 $\pm$ 0.03 &  0.172 $\pm$ 0.019 & 0.14 \\
 \\
 123352 &  42.16333 & 23.27444 & 18.65 $\pm$ 0.08 &  0.599 $\pm$ 0.040 & 0.57 \\
 124056 &  44.41000 & 23.80333 & 20.82 $\pm$ 0.29 &  (0.030 $\pm$ 0.010) & 0.37 \\
 124629 &  44.02333 &  2.80861 & 19.53 $\pm$ 0.20 &  (0.030 $\pm$ 0.010) & 0.25 \\
 124635 &  42.15958 & 19.32806 & 18.00 $\pm$ 0.05 &  3.519 $\pm$ 0.080 & 0.26 \\
 171459 & 116.18209 & 25.14056 & 17.29 $\pm$ 0.03 &  0.502 $\pm$ 0.020 & 0.09 \\
 174585 & 114.04292 &  9.98639 & 17.38 $\pm$ 0.04 &  0.821 $\pm$ 0.028 & 0.09 \\
 174605 & 117.59000 &  7.79444 & 17.17 $\pm$ 0.04 &  0.419 $\pm$ 0.011 & 0.05 \\
 182595 & 132.80042 & 27.88000 & 16.17 $\pm$ 0.04 &  2.271 $\pm$ 0.024 & 0.10 \\
 191706 & 142.55333 & 19.99056 & 16.67 $\pm$ 0.03 &  (0.030 $\pm$ 0.010) & 0.10 \\
 191791 & 137.22417 & 14.58389 & 17.34 $\pm$ 0.02 &  (0.030 $\pm$ 0.010) & 0.09 \\
\enddata
\tablecomments{H$\alpha$ fluxes enclosed in parentheses represent upper limits.   See text for details.}
\tablecomments{Table 1 is published in its entirety in the machine-readable format.  A portion is shown here for guidance regarding its form and content.}
\end{deluxetable*}

\subsection{Measuring H$\alpha$ Fluxes from WIYN Images}\label{sec:flux}

Measurements of both the H$\alpha$ and R-band fluxes were carried out on our fully processed images using a photometry script which allows the user to first mask any intervening objects and then to interactively determine a suitable aperture size that includes the total flux from the object.   The interactive photometry script used here is very similar to the one described in \citet{gormanous26}, which was used to measure the UV fluxes for the SHIELD galaxies from GALEX survey images.

The measured instrumental fluxes were next converted to standardized magnitudes and calibrated H$\alpha$ fluxes.  The measurements were corrected for atmospheric extinction and calibrated using zero-point constants derived from our standard star observations.  The calibrated H$\alpha$ fluxes were then corrected for bandpass effects of our narrowband and broadband filters which depend on the precise redshift of the galaxy.  This correction also accounts for the small amount of H$\alpha$ flux removed during the continuum subtraction due to the presence of H$\alpha$ emission in the R-band filter.  Finally, we apply a statistical correction for the presence of [\ion{N}{2}] line emission that falls within the narrowband image.  Our correction methods follow those used by \citet{kennicutt2008} and \citet{vansistine2016}.  Specifically, we fit a relationship between log([\ion{N}{2}]/H$\alpha$) and R-band absolute magnitude for a sample of galaxies where these values are available.  We then use the resulting relationship in conjunction with the measured value of M$_R$ to estimate log([\ion{N}{2}]/H$\alpha$) for each SHIELD galaxy.  We note that since the luminosities of the SHIELD galaxies are all quite low (see Figure~\ref{fig:histograms}b below), the typical [\ion{N}{2}] correction is in all cases quite small (less than 3\% of the H$\alpha$ flux).

Our measured R-band magnitudes and fully corrected H$\alpha$ fluxes are presented in Table~\ref{tab:haflux}.  Column 1 gives the galaxy name (AGC number) and columns 2 and 3 list the celestial coordinates for each source.  Column 4 gives the apparent R-band magnitude derived from our direct measurements of the R-band images.  Given the short exposure times and small telescope aperture, the formal errors of our apparent magnitudes tend to be relatively large, particularly for the low surface brightness SHIELD galaxies.  Column 5 lists the calibrated and corrected H$\alpha$ fluxes for all of the detected sources, in units of erg s$^{-1}$ cm$^{-2}$, while Column 6 provides the Galactic absorption value at the location of each galaxy from \citet{schlafly2011}.  This latter value is used in our Galactic absorption correction described in the next section.

Our observations resulted in measurable H$\alpha$ fluxes for 55 out of 82 SHIELD galaxies (67.1\%). The remaining galaxies had no detectable H$\alpha$ flux at the depth of our observations.  For the non-detections we assign a conservative upper limit for their H$\alpha$ fluxes by taking the average flux uncertainties for the eight SHIELD galaxies with the weakest {\it detected} fluxes ($\langle\sigma\rangle$ = 1.0 $\times$ 10$^{-16}$ erg s$^{-1}$ cm$^{-2}$) and multiply this by 3.0.   Non-detections are indicated in Table~\ref{tab:haflux} by parentheses around their 3$\sigma$ upper limit values.

\subsection{H$\alpha$ SFRs for SHIELD Galaxies}\label{sec:sfr}

H$\alpha$-based SFRs were calculated for the SHIELD galaxies using the conversion relation derived by  \citet{Kennicutt1998a}:
\begin{equation}
 SFR_{H\alpha}=7.9  \times 10^{-42} \cdot L_{H\alpha}, 
\end{equation}
where SFR$_{H\alpha}$ has units $M_\odot$ yr$^{-1}$ and the H$\alpha$ luminosity L$_{H\alpha}$ has units of ergs s$^{-1}$.  This SFR conversion factor is derived under the assumption of stellar populations with solar metallicity undergoing continuous star formation for at least 10 Myr, and using a Salpeter IMF \citep{salpeter1955} integrated over the mass range of 0.1 to 100 M$_\odot$.  It is worth stressing that any deviations from these assumptions will negate the validity of the conversion from H$\alpha$ luminosity to the SFR.

The H$\alpha$ fluxes were used to determine the H$\alpha$ luminosities with the following equation:
\begin{equation}
  L_{H\alpha}=4\pi D^2 \times F_{H\alpha} \times 10^{0.4A_{H\alpha}},
\end{equation}
where D is the distance in cm, $F_{H\alpha}$ is the H$\alpha$ flux  in ergs s$^{-1}$ cm$^{-2}$, and A$_{H\alpha}$ is the Galactic absorption at H$\alpha$ in magnitudes. For this calculation we adopted the simplifying approximation that A$_{H\alpha}$ = A$_{R}$.   Both $F_{H\alpha}$ and A$_{H\alpha}$ are listed in Table~\ref{tab:haflux}.  We note that this correction is for Galactic absorption only.  No corrections for absorption internal to the individual galaxies are applied in this study.  Given the extremely low-mass (and thus, low-metallicity) nature of the SHIELD sample we do not expect the internal absorption to be large for most of our sources. 
For the 27 SHIELD galaxies with no H$\alpha$ detections we computed L$_{H\alpha}$ and SFR$_{H\alpha}$ using the 3$\sigma$ upper limit flux values.

As described in previous papers presenting results on the SHIELD galaxies \citep[e.g.,][]{mcquinn2021, gormanous26}, distances for the SHIELD galaxies have been determined using one of two methods. The more precise method involves using TRGB distances from \citet{mcquinn2014} and \citet{mcquinn2021} when possible.  This method uses the resolved stellar photometry from HST images to create a color-magnitude diagram (CMD) for each galaxy.  The CMDs are used to determine the apparent magnitudes of the TRGB which can then be used with the known TRGB absolute magnitude to determine distances.   TRGB distances are currently available for 33 of the 82 SHIELD galaxies.  The second source of distances are flow-model distances. This process uses the measured line-of-sight velocity of the galaxy combined with Hubble’s Law and a flow model (to correct for any local gravitational effects) to derive the distance. This method is substantially less accurate than the TRGB method \citep{mcquinn2014, mcquinn2021}. Flow model distances for all ALFALFA detected galaxies are derived and included in the ALFALFA survey tables \citep{haynes2011, haynes2018}, based on the flow model of \citet{masters_2005}. The distances as well as the H$\alpha$ luminosities and SFRs for the SHIELD galaxies are given in Table \ref{tab:derived}.


\startlongtable
\centerwidetable
\begin{longrotatetable}
\begin{deluxetable*}{crcccccccccc}
\digitalasset
\tabletypesize{\scriptsize}
\tablenum{2}
\tablecaption{Derived Properties of the SHIELD Galaxies\label{tab:derived}}
\tablehead{
AGC  \# & Distance\ \ \ \ & Source & M$_R$  & log(M$_{star}$)  & log(M$_{HI}$)  & log(M$_{baryonic}$) &  log(L$_{H\alpha}$) & log(SFR$_{H\alpha}$)  & log(SFR$_{H\alpha}$/SFR$_{FUV}$) & log(sSFR$_{H\alpha}$)  & M$_{gas}$/M$_{baryonic}$ \\  
  & Mpc\ \ \ \ \ \ \  &  & mag  & M$_\odot$  & M$_\odot$  & M$_\odot$  & erg/s &  M$_\odot$ yr$^{-1}$  &   & yr$^{-1}$  \\ 
(1) & (2)\ \ \ \ \ \ \ \ & (3) & (4) & (5) & (6) & (7) & (8) & (9) & (10) & (11) & (12)  
}
 
\startdata 
 102728 & 12.41 $\pm$ 0.64 &  1 & -11.60 $\pm$ 0.14 &  6.32 $\pm$ 0.20 &   7.05 $\pm$ 0.06 &   7.24 $\pm$ 0.06 &  (36.79 $\pm$ 0.15) & (-4.31 $\pm$ 0.16) & (-1.37 $\pm$ 0.21) & (-10.63 $\pm$ 0.25) &  0.88 $\pm$ 0.17 \\
 103722 &  5.64 $\pm$ 0.15 &  3 & -11.73 $\pm$ 0.07 &  6.43 $\pm$ 0.12 &   7.13 $\pm$ 0.03 &   7.32 $\pm$ 0.03 &  37.33 $\pm$ 0.04 & -3.78 $\pm$ 0.06 & -0.89 $\pm$ 0.15 & -10.21 $\pm$ 0.13 &  0.87 $\pm$ 0.07 \\
 104208 &  9.70 $\pm$ 0.97 &  2 & -10.61 $\pm$ 0.28 &  5.99 $\pm$ 0.18 &   6.95 $\pm$ 0.10 &   7.11 $\pm$ 0.09 &  (36.63 $\pm$ 0.17) & (-4.47 $\pm$ 0.17) & (-1.39 $\pm$ 0.26) & (-10.46 $\pm$ 0.25) &  0.92 $\pm$ 0.28 \\
 110482 &  7.82 $\pm$ 0.21 &  1 & -14.03 $\pm$ 0.07 &  7.39 $\pm$ 0.12 &   7.28 $\pm$ 0.03 &   7.70 $\pm$ 0.06 &  38.23 $\pm$ 0.03 & -2.87 $\pm$ 0.05 & -0.41 $\pm$ 0.14 & -10.26 $\pm$ 0.13 &  0.51 $\pm$ 0.08 \\
 111164 &  5.11 $\pm$ 0.07 &  1 & -12.06 $\pm$ 0.04 &  6.57 $\pm$ 0.12 &   6.60 $\pm$ 0.03 &   6.96 $\pm$ 0.05 &  37.62 $\pm$ 0.01 & -3.48 $\pm$ 0.05 & -0.39 $\pm$ 0.14 & -10.05 $\pm$ 0.13 &  0.59 $\pm$ 0.08 \\
 111946 &  9.02 $\pm$ 0.29 &  1 & -12.70 $\pm$ 0.08 &  6.68 $\pm$ 0.12 &   7.16 $\pm$ 0.03 &   7.39 $\pm$ 0.04 &  37.71 $\pm$ 0.03 & -3.39 $\pm$ 0.06 & -0.82 $\pm$ 0.14 & -10.07 $\pm$ 0.13 &  0.80 $\pm$ 0.09 \\
 111977 &  5.96 $\pm$ 0.11 &  1 & -13.61 $\pm$ 0.05 &  7.10 $\pm$ 0.12 &   6.85 $\pm$ 0.03 &   7.35 $\pm$ 0.07 &  37.83 $\pm$ 0.02 & -3.27 $\pm$ 0.05 & -0.56 $\pm$ 0.14 & -10.37 $\pm$ 0.13 &  0.43 $\pm$ 0.08 \\
 112503 & 10.20 $\pm$ 1.02 &  2 & -13.44 $\pm$ 0.22 &  7.31 $\pm$ 0.17 &   7.14 $\pm$ 0.10 &   7.59 $\pm$ 0.10 &  38.51 $\pm$ 0.09 & -2.60 $\pm$ 0.10 &   ...   &  -9.91 $\pm$ 0.20 &  0.48 $\pm$ 0.15 \\
 112505 & 10.30 $\pm$ 1.03 &  2 &  -9.53 $\pm$ 0.41 &  6.10 $\pm$ 0.17 &   7.12 $\pm$ 0.09 &   7.28 $\pm$ 0.09 &  (36.64 $\pm$ 0.17) & (-4.46 $\pm$ 0.17) & (-1.48 $\pm$ 0.24) & (-10.56 $\pm$ 0.24) &  0.93 $\pm$ 0.27 \\
 112521 &  6.58 $\pm$ 0.18 &  1 & -11.61 $\pm$ 0.07 &  6.25 $\pm$ 0.12 &   6.85 $\pm$ 0.03 &   7.05 $\pm$ 0.03 &  37.01 $\pm$ 0.05 & -4.09 $\pm$ 0.07 & -0.73 $\pm$ 0.15 & -10.34 $\pm$ 0.14 &  0.84 $\pm$ 0.10 \\
\\
 123352 &  9.72 $\pm$ 0.25 &  1 & -11.85 $\pm$ 0.09 &  6.57 $\pm$ 0.12 &   7.18 $\pm$ 0.03 &   7.38 $\pm$ 0.03 &  38.08 $\pm$ 0.04 & -3.03 $\pm$ 0.06 & -0.80 $\pm$ 0.16 &  -9.60 $\pm$ 0.13 &  0.85 $\pm$ 0.08 \\
 124056 &  5.90 $\pm$ 0.59 &  2 &  -8.40 $\pm$ 0.37 &  6.42 $\pm$ 0.17 &   6.55 $\pm$ 0.10 &   6.87 $\pm$ 0.09 &  (36.26 $\pm$ 0.17) & (-4.84 $\pm$ 0.17) &   ...   & (-11.26 $\pm$ 0.24) &  0.64 $\pm$ 0.19 \\
 124629 & 10.60 $\pm$ 1.06 &  2 & -10.85 $\pm$ 0.30 &  6.21 $\pm$ 0.17 &   7.06 $\pm$ 0.10 &   7.23 $\pm$ 0.09 &  (36.72 $\pm$ 0.17) & (-4.39 $\pm$ 0.17) & (-1.44 $\pm$ 0.26) & (-10.60 $\pm$ 0.24) &  0.90 $\pm$ 0.27 \\
 124635 &  7.90 $\pm$ 0.79 &  2 & -11.75 $\pm$ 0.22 &  7.19 $\pm$ 0.17 &   7.05 $\pm$ 0.09 &   7.49 $\pm$ 0.10 &  38.53 $\pm$ 0.09 & -2.57 $\pm$ 0.10 & -0.02 $\pm$ 0.19 &  -9.76 $\pm$ 0.20 &  0.50 $\pm$ 0.15 \\
 171459 & 11.80 $\pm$ 1.18 &  2 & -13.17 $\pm$ 0.22 &  6.94 $\pm$ 0.17 &   7.21 $\pm$ 0.09 &   7.48 $\pm$ 0.08 &  37.96 $\pm$ 0.09 & -3.14 $\pm$ 0.10 & -0.45 $\pm$ 0.19 & -10.08 $\pm$ 0.20 &  0.71 $\pm$ 0.21 \\
 174585 &  7.89 $\pm$ 0.21 &  1 & -12.19 $\pm$ 0.07 &  6.56 $\pm$ 0.12 &   6.90 $\pm$ 0.04 &   7.16 $\pm$ 0.04 &  37.82 $\pm$ 0.03 & -3.28 $\pm$ 0.05 & -0.34 $\pm$ 0.14 &  -9.84 $\pm$ 0.13 &  0.75 $\pm$ 0.10 \\
 174605 & 10.89 $\pm$ 0.28 &  1 & -13.07 $\pm$ 0.07 &  6.94 $\pm$ 0.12 &   7.27 $\pm$ 0.03 &   7.53 $\pm$ 0.04 &  37.80 $\pm$ 0.03 & -3.31 $\pm$ 0.05 & -0.68 $\pm$ 0.15 & -10.25 $\pm$ 0.13 &  0.74 $\pm$ 0.09 \\
 182595 &  9.02 $\pm$ 0.28 &  1 & -13.70 $\pm$ 0.08 &  7.22 $\pm$ 0.12 &   6.91 $\pm$ 0.04 &   7.44 $\pm$ 0.07 &  38.39 $\pm$ 0.03 & -2.72 $\pm$ 0.05 &  0.06 $\pm$ 0.14 &  -9.94 $\pm$ 0.13 &  0.40 $\pm$ 0.08 \\
 191706 &  8.20 $\pm$ 0.82 &  2 & -12.99 $\pm$ 0.22 &  7.08 $\pm$ 0.17 &   7.20 $\pm$ 0.09 &   7.52 $\pm$ 0.08 &  (36.42 $\pm$ 0.17) & (-4.68 $\pm$ 0.17) & (-1.90 $\pm$ 0.24) & (-11.76 $\pm$ 0.24) &  0.64 $\pm$ 0.18 \\
 191791 &  9.50 $\pm$ 0.95 &  2 & -12.64 $\pm$ 0.22 &  6.77 $\pm$ 0.17 &   6.78 $\pm$ 0.16 &   7.14 $\pm$ 0.12 &  (36.55 $\pm$ 0.17) & (-4.55 $\pm$ 0.17) & (-1.24 $\pm$ 0.24) & (-11.32 $\pm$ 0.24) &  0.58 $\pm$ 0.27 \\
\enddata 

\tablecomments{Source of distances (Column 3): 1 = distance based on TRGB method, taken from \citet{mcquinn2014} and \citet{mcquinn2021}; 2 = distance based on flow model of \citet{masters_2005}, taken from AFLALFA catalogs \citep{haynes2011, haynes2018}; 3 = distance derived using TRGB method, based on unpublished analysis (K. McQuinn, private communication).}
\tablecomments{H$\alpha$-related quantities in Column 8-11 that are enclosed in parentheses represent upper limits.}
\tablecomments{Galaxies with no value given in Column 10 have no FUV flux measurement (see Paper 1)}
\tablecomments{Table 2 is published in its entirety in the machine-readable format.  A portion is shown here for guidance regarding its form and content.}
\end{deluxetable*}
\end{longrotatetable}


\section{Results}\label{sec:results}

\subsection{Properties of the SHIELD Galaxies}\label{sec:prop}

We present the derived properties of the 82 SHIELD galaxies in Table~\ref{tab:derived}.  Column 1 lists the AGC number, column 2 presents the distance (in Mpc) used to derive distance-dependent quantities, and column 3 gives the source of these distances, as specified in the Table notes. Column 4 lists the R-band absolute magnitude for each galaxy.  These have been corrected for Galactic absorption using the value of A$_R$ for each galaxy from \citet{schlafly2011}.  Columns 5-7 present the stellar masses, \ion{H}{1} masses, and baryonic masses for the SHIELD galaxies.  Stellar masses are taken from \citet{marine2023} and are derived using measurements of the 3.6 $\mu$m NIR fluxes measured from Spitzer images.  The \ion{H}{1} masses are computed from ALFALFA-based 21-cm flux measurements \citep{haynes2011, haynes2018} but make use of our updated distances when available.  The baryonic masses are the sum of the stellar and gas masses, where the latter are computed as M$_{gas}$ = 1.35$\cdot$M$_{HI}$, where the factor of 1.35 accounts for the amount of helium present in the ISM of these galaxies.  We do not include an estimate of the mass of ionized gas in our values of M$_{gas}$ since a simple calculation suggests that the ionized gas will be a negligible fraction of the total gas mass (less than 0.1\%).  Furthermore, because the CO molecule is notoriously difficult to detect in low-metallicity dwarf galaxies \citep[e.g.,][]{sage1992, Taylor_1998, leroy2005, schruba2012, bolatto2013, warren2015}, we include no estimates of the molecular gas mass in our values for M$_{gas}$.

Columns 8 and 9 present the H$\alpha$ luminosities and SFRs for the SHIELD galaxies, derived as described in Section~\ref{sec:sfr}.  For objects that were not detected in our H$\alpha$ NB images, the values listed here are upper limits based on the 3$\sigma$ upper limit fluxes described in \S~\ref{sec:flux} and are enclosed in parentheses.  Column 10 lists the logarithm of the ratio of the H$\alpha$ SFR to the FUV SFR \citep{gormanous26}, while Column 11 gives the logarithm of the specific H$\alpha$ star-formation rate (sSFR$_{H\alpha}$ = SFR$_{H\alpha}$/M$_{stellar}$).  Again, objects in parentheses represent upper limits for both of these quantities.  Finally, Column 12 presents the gas mass fraction M$_{gas}$/M$_{baryonic}$.  Following \citet{gormanous26}, we feel that it is more appropriate to present the gas mass fraction relative to the baryonic mass rather than the stellar mass, since the SHIELD galaxies tend to be gas-mass dominated (M$_{gas}$ $>$ M$_{star}$), with several galaxies having M$_{gas}$/M$_{star}$ $>$ 10.

Since the current paper makes use of FUV SFRs presented in \citet{gormanous26}, it is relevant to point out that these SFRs were computed using the conversion factor derived by \citet{mcquinn2015b}.  The \citet{mcquinn2015b} analysis uses a Salpeter IMF, making it consistent with the H$\alpha$ conversion of \citet{kennicutt2008} (e.g., Equation 1 above).  However, they use a slightly different mass range of 0.1 to 120 M$_\odot$.  \citet{mcquinn2015b} argue that the different mass ranges will have only a small impact on the resulting conversion factor (they estimate a $\sim$5\% difference).

\begin{figure*}
\centering

\includegraphics[width=5.70in]{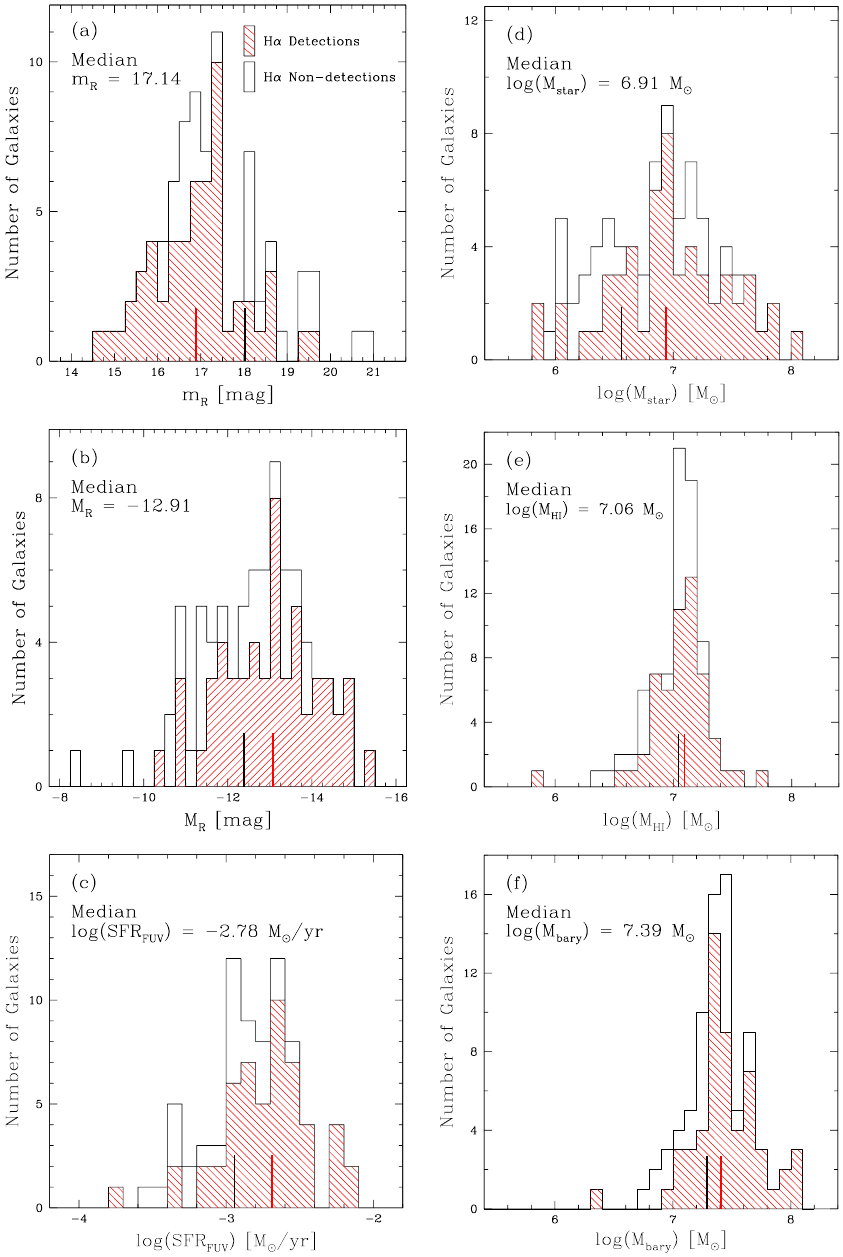}

\caption{Histograms showing the properties of the full sample of SHIELD galaxies (N=82).  The panels in the left column present the distributions of (a) apparent R magnitude, (b) absolute R magnitude, and (c) FUV SFR. The panels in the right include histograms of (d) stellar, (e) \ion{H}{1}, and (f) baryonic masses.  In all histograms, the H$\alpha$-detected SHIELD galaxies are delimited by the red cross hatching, and the red vertical bar shows the median value for that subsample. The black histograms show the full sample, while the black vertical bar denotes the median of the H$\alpha$ non-detections.}
\label{fig:histograms}
\end{figure*}

Because the current paper and its companion paper \citep{gormanous26} are the first to present the full catalog of 82 SHIELD galaxies, it is worthwhile to illustrate the properties of the galaxies that make up this sample.  Accordingly, Figure~\ref{fig:histograms} presents a series of histograms of several relevant quantities.  

Panel (a) of Figure~\ref{fig:histograms} plots the distribution of R-band apparent magnitudes for the SHIELD galaxies.  As described in Section~\ref{sec:flux}, these R magnitudes are measured from our own broadband images.  The SHIELD galaxies exhibit a broad range of brightnesses, from R = 14.5 to 20.8.  The median R magnitude is 17.14.  The red cross-hatched portion of the histogram designates the H$\alpha$-detected galaxies, while the unshaded portion of the histogram are the H$\alpha$ non-detections.  Interestingly, the non-detections are distributed over nearly the full range of the apparent magnitudes found in SHIELD, but with a tendency for the fainter galaxies to be among the non-detections.  The median R-band magnitude for the H$\alpha$ detections is 16.89, while the corresponding value for the non-detections is R = 18.02.

Figure~\ref{fig:histograms}, panel(b) shows the distribution of R-band absolute magnitudes. The SHIELD galaxies have median M$_R$ of $-$12.91 with values covering the range $-$8.4 to $-$15.4.  There is only a modest difference between the H$\alpha$ detections and non-detections, with the latter tending to be somewhat lower luminosities (median M$_R$ = $-$12.39) compared to the former (median M$_R$ = $-$13.08).  Overall, the SHIELD galaxies are seen to be confined to quite low luminosities.  By way of comparison, the SMC has M$_R$ $\sim$ $-$16.9.  Hence, the median SHIELD galaxy is $\sim$4.0 magnitudes (factor of $\sim$40.0) lower luminosity.

Panel (c) presents the distribution of FUV SFRs.  Despite the fact that the current paper focuses on the H$\alpha$ measurements of the SHIELD galaxies, we feel that it is more relevant to present the FUV SFRs as opposed to H$\alpha$ SFRs when considering the overall properties of the SHIELD sample.  As emphasized in \citet{gormanous26} as well as \S~\ref{sec:HAsfr} of the current paper, the instantaneous SFR, as measured by the H$\alpha$ luminosity, deviates from the SFR$_{FUV}$ for log(SFR$_{H\alpha}$) below $-$2.5.  Hence, for the majority of the SHIELD galaxies, SFR$_{H\alpha}$ will be an underestimate of their recent average SFR indicated by FUV measurements, often by large factors.  Conversely, \citet{gormanous26} showed that the FUV measurements provide a robust estimate of the SFRs for dwarf galaxies down to the lowest masses.  For example, Figure 7 in \citet{gormanous26} shows a well-defined linear relationships between log(M$_{Stellar}$) and FUV SFR over the mass range of 10$^6$ - 10$^{11}$ M$_\odot$, while Figure 8 from the same study shows a similar result for log(M$_{Baryonic}$) over a comparable mass range.
Because only 75 of the 82 SHIELD galaxies were observed by GALEX, histogram (c) is the only one of the panels in Figure~\ref{fig:histograms} where the {\it full} SHIELD sample is not plotted.

The median value for log(SFR$_{FUV}$) is $-$2.78.  The corresponding median value for log(SFR$_{H\alpha}$) is $-$3.59 ($-$3.24 if only H$\alpha$ detections are considered), where the large difference reflects the fact that the instantaneous SFR does not track the time-averaged SFR at low SFR values, as described above. The key take-away point of the figure is the extremely low SFRs represented by the SHIELD galaxies. The SHIELD object with the {\it highest} SFR (log(SFR$_{FUV}$) = $-$2.15) is only $\sim$0.24-0.71\% the SFR of the Milky Way \citep[1-3 M$_\odot$ yr$^{-1}$;][]{chomiuk2011, licquia2015, elia2022}, emphasizing the extreme nature of the sample.  

The righthand column of Figure~\ref{fig:histograms} plots the distributions of three different mass indicators for the SHIELD galaxies: stellar mass (panel (d)), \ion{H}{1} mass (panel (e)), and baryonic mass (panel (f)).  All masses are given in units of M$_\odot$.  Similar to the panels discussed previously, the H$\alpha$ non-detections are fairly evenly distributed in mass relative to the full sample in all three mass plots, with only a slight trend for the non-detections to be found among the lower-mass systems.  The median masses for each distribution are indicated in the plots.  We note that the median \ion{H}{1} mass is higher than the median stellar mass, consistent with the point stressed in \citet{gormanous26} that the SHIELD galaxies are gas dominated.

The stellar masses show a much broader distribution than do either the \ion{H}{1} or baryonic masses.  The limited \ion{H}{1} mass range exhibited by the SHIELD galaxies in panel (e) is an artifact of the selection function used to create the sample.  The original SHIELD sample was selected from the overall ALFALFA survey catalogs \citep{giovanelli2005, haynes2011, haynes2018} with the requirement that log(M$_{HI}$) $\le$ 7.20.  However, the original distance estimates of many of the SHIELD galaxies proved to be underestimated \citep{mcquinn2021}, and with our improved TRGB distances 15 of the 82 SHIELD galaxies are found to have \ion{H}{1} masses slightly above this mass limit.  Only three galaxies have log(M$_{HI}$) $>$ 7.4.

The distribution of baryonic masses shown in panel (f) of Figure~\ref{fig:histograms} is also quite narrow, reflecting the shape of the \ion{H}{1} mass distribution in the panel above.  This is as expected, given that the SHIELD galaxies are gas dominated.  The median value of M$_{baryonic}$ is 2.48 $\times$ 10$^7$ M$_\odot$, and 84\% of the sample (69 of 82) have masses between log(M$_{baryonic}$) of 7.0 and 7.8.

\subsection{H$\alpha$ Images and Morphologies}\label{sec:images}

\begin{figure*}
\centering
\includegraphics[width=7.1in]{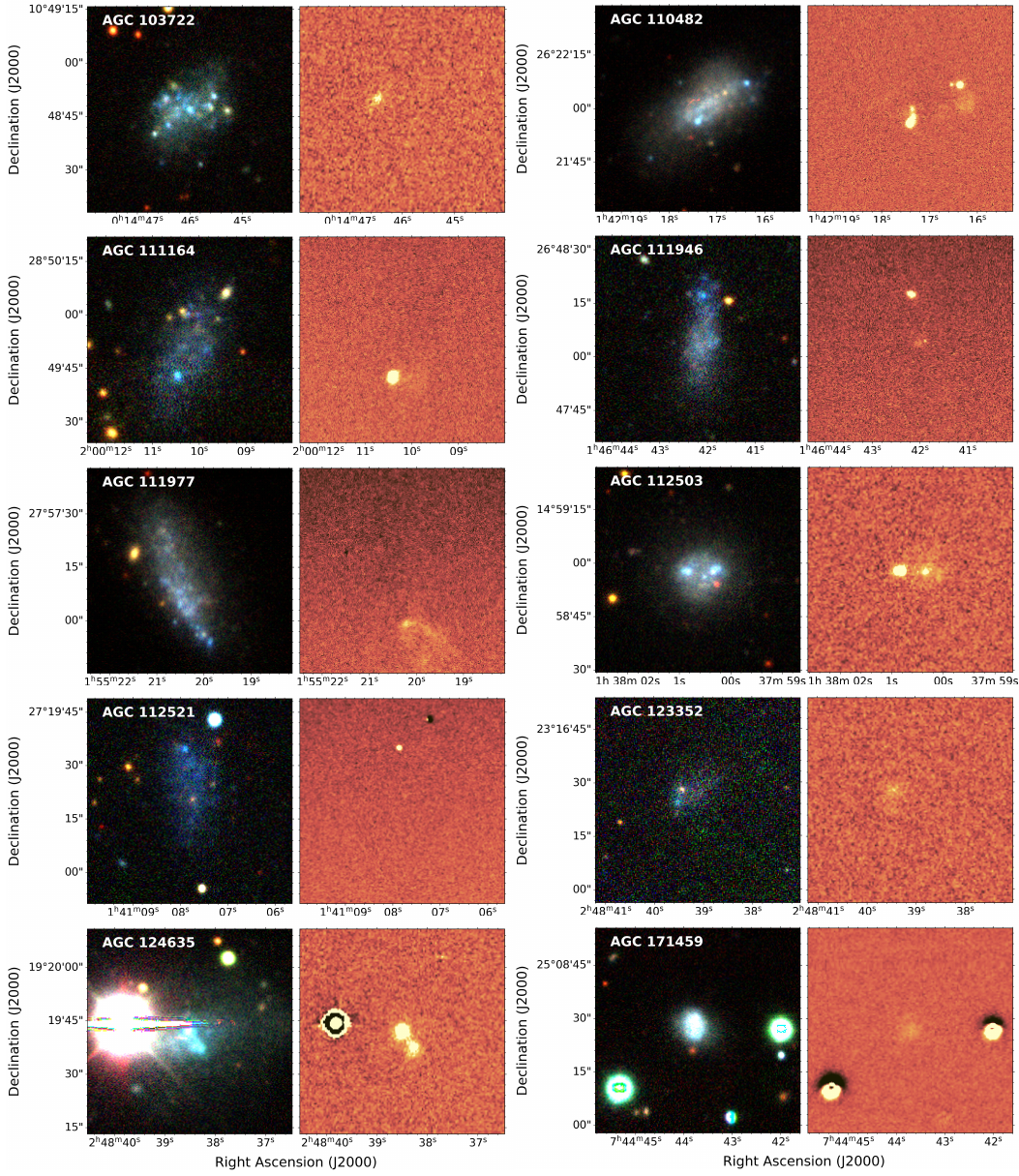}
\caption{ Images of the H$\alpha$-detected SHIELD galaxies.  For each galaxy we show a pair of images, consisting of a color composite image created using {\it giz} DESI Legacy imaging data (left) and the continuum-subtracted H$\alpha$ narrowband image (right).  Each image segment covers an area of roughly 56 $\times$ 56 arcsec, and in all cases N is up and E is left.  The broadband image of AGC 124635 (left column, 5th row) is impacted by the bleed trail from the nearby star, although the two knots of high surface brightness emission associated with the two \ion{H}{2} regions are visible.}
\label{fig:SHIELDimages1}
\end{figure*}

\begin{figure*}
\centering
\includegraphics[width=7.1in]{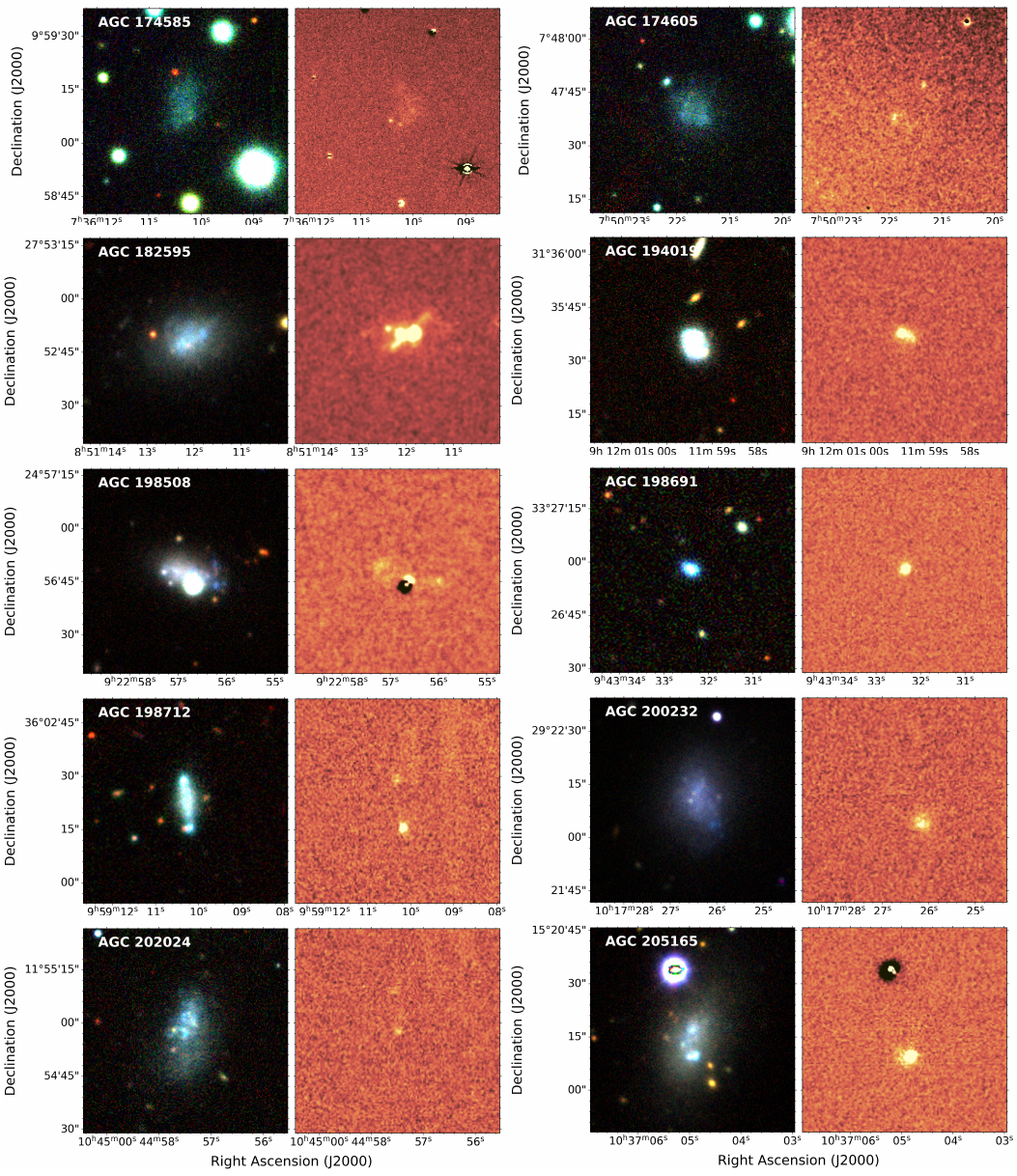}
\caption{ Images of the H$\alpha$-detected SHIELD galaxies.  For each galaxy we show a pair of images, consisting of a color composite image created using {\it giz} DESI Legacy imaging data (left) and the continuum-subtracted H$\alpha$ narrowband image (right).  Each image segment covers an area of roughly 56 $\times$ 56 arcsec, and in all cases N is up and E is left.}
\label{fig:SHIELDimages2}
\end{figure*}

\begin{figure*}
\centering
\includegraphics[width=7.1in]{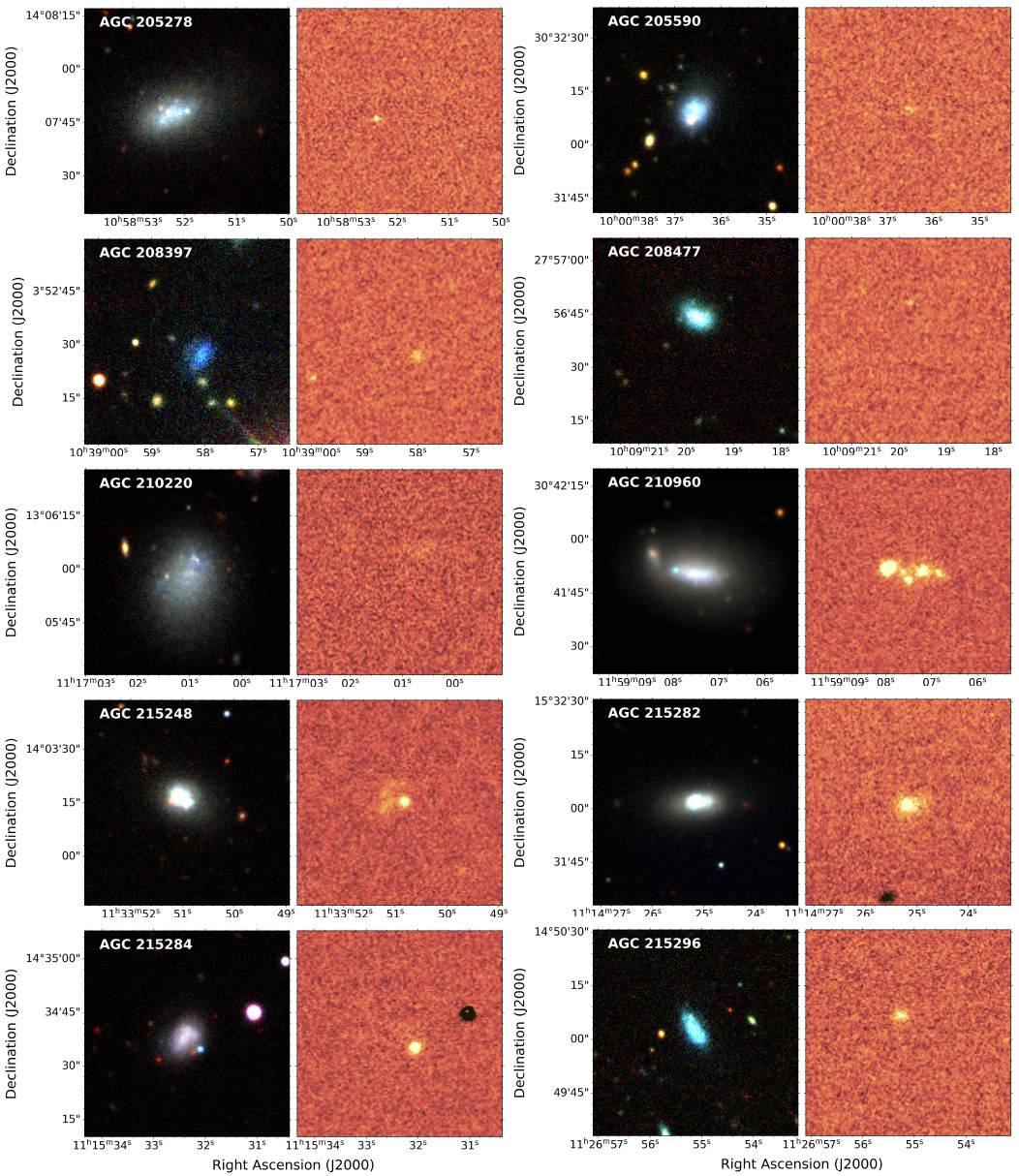}
\caption{ Images of the H$\alpha$-detected SHIELD galaxies.  For each galaxy we show a pair of images, consisting of a color composite image created using {\it giz} DESI Legacy imaging data (left) and the continuum-subtracted H$\alpha$ narrowband image (right).  Each image segment covers an area of roughly 56 $\times$ 56 arcsec, and in all cases N is up and E is left.}
\label{fig:SHIELDimages3}
\end{figure*}

\begin{figure*}
\centering
\includegraphics[width=7.1in]{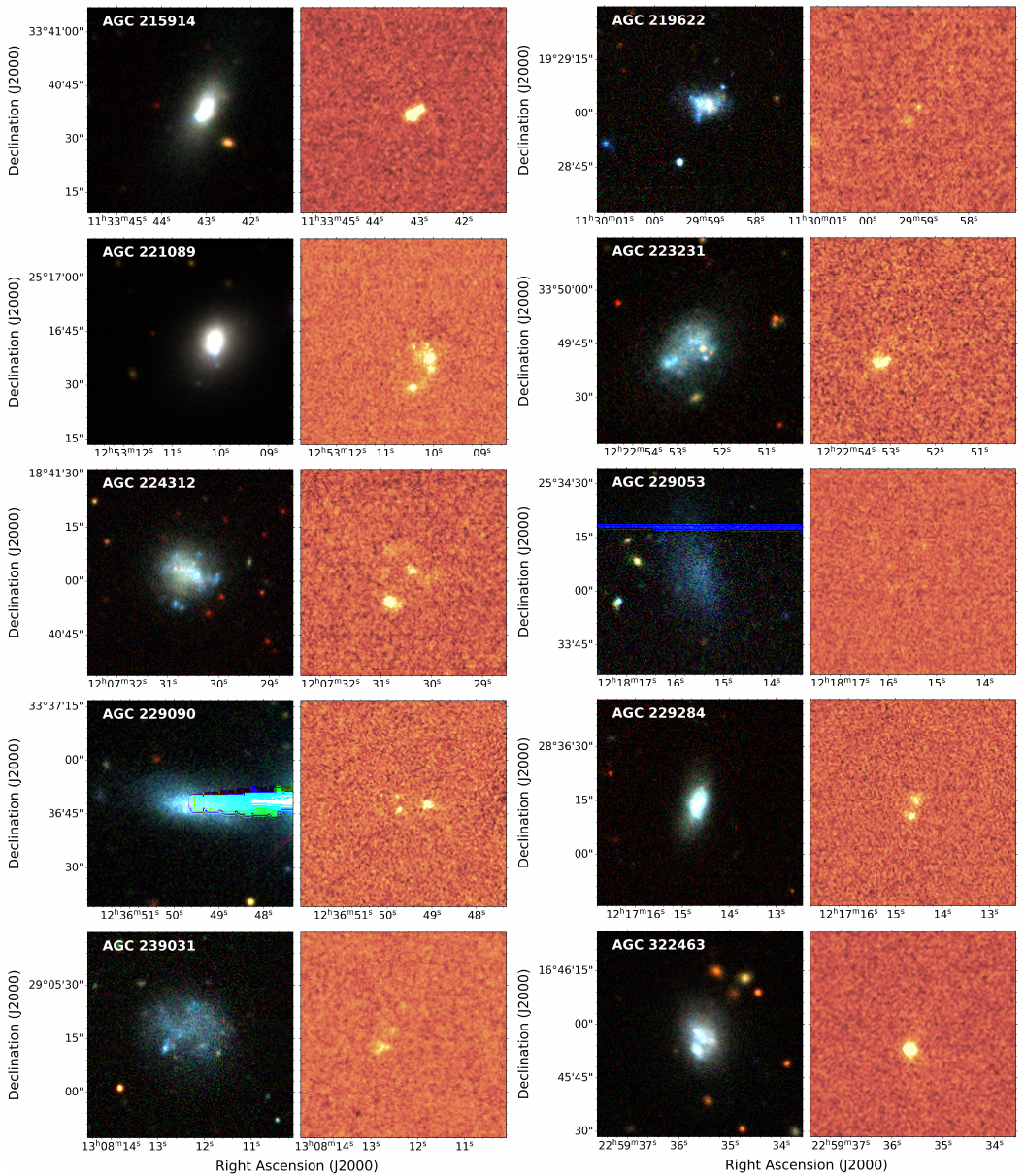}
\caption{ Images of the H$\alpha$-detected SHIELD galaxies.  For each galaxy we show a pair of images, consisting of a color composite image created using {\it giz} DESI Legacy imaging data (left) and the continuum-subtracted H$\alpha$ narrowband image (right).  Each image segment covers an area of roughly 56 $\times$ 56 arcsec, and in all cases N is up and E is left.  Note that in the broadband image of AGC 229090 (left column, 4th row) the galaxy is largely obscured by a bleed trail from a bright star located just to the west of the galaxy.}
\label{fig:SHIELDimages4}
\end{figure*}

\begin{figure*}
\centering
\includegraphics[width=7.1in]{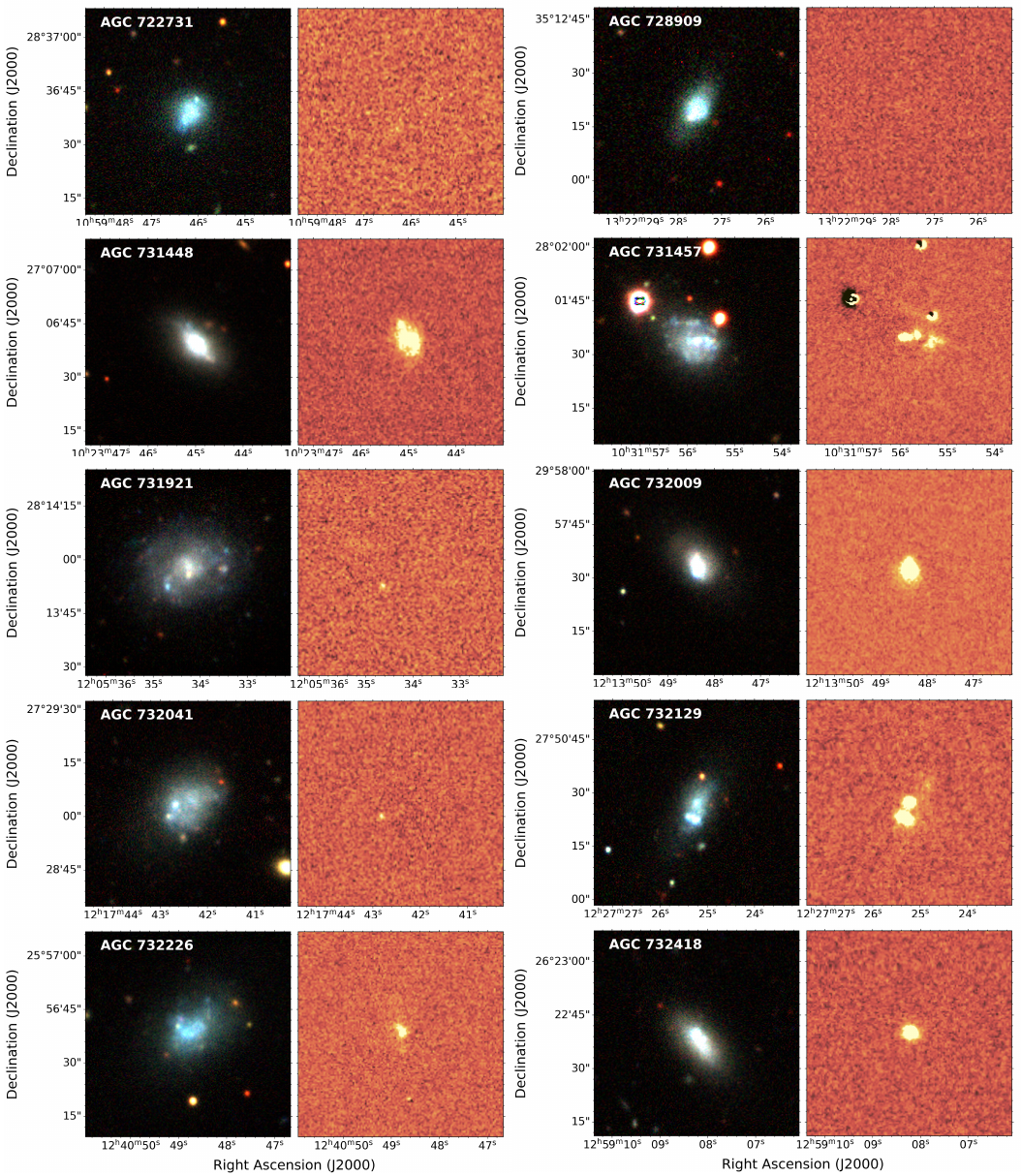}
\caption{ Images of the H$\alpha$-detected SHIELD galaxies.  For each galaxy we show a pair of images, consisting of a color composite image created using {\it giz} DESI Legacy imaging data (left) and the continuum-subtracted H$\alpha$ narrowband image (right).  Each image segment covers an area of roughly 56 $\times$ 56 arcsec, and in all cases N is up and E is left.}
\label{fig:SHIELDimages5}
\end{figure*}

\begin{figure*}
\centering
\includegraphics[width=7.1in]{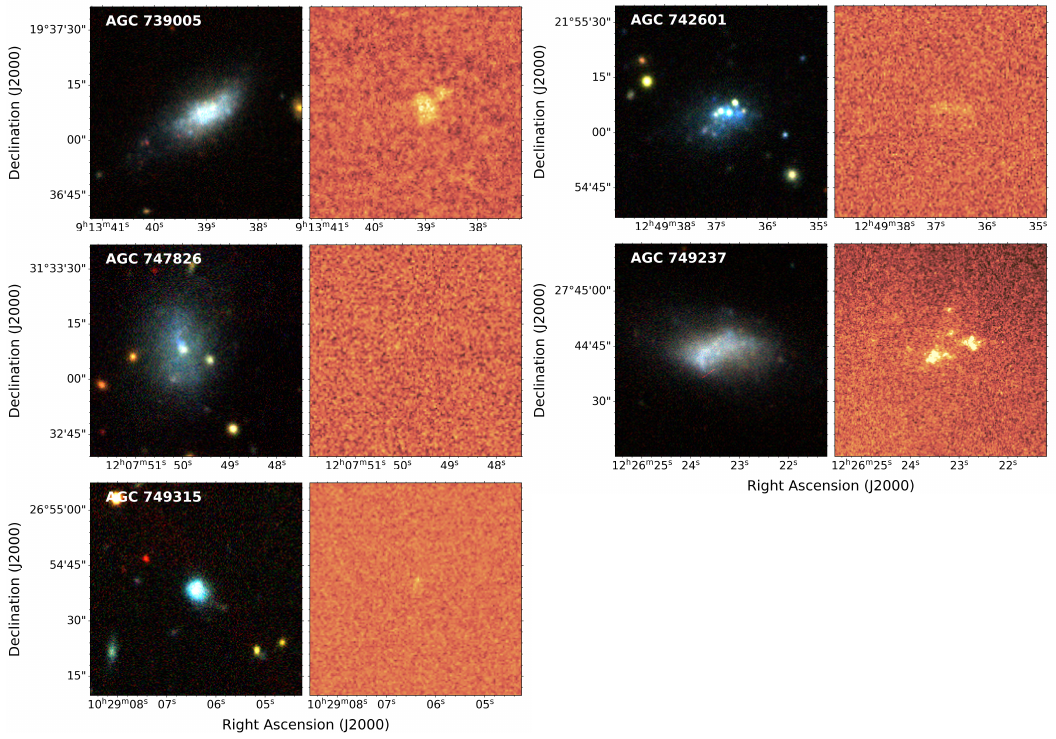}
\caption{ Images of the H$\alpha$-detected SHIELD galaxies.  For each galaxy we show a pair of images, consisting of a color composite image created using {\it giz} DESI Legacy imaging data (left) and the continuum-subtracted H$\alpha$ narrowband image (right).  Each image segment covers an area of roughly 56 $\times$ 56 arcsec, and in all cases N is up and E is left.}
\label{fig:SHIELDimages6}
\end{figure*}

\begin{figure*}
\centering
\includegraphics[width=7.1in]{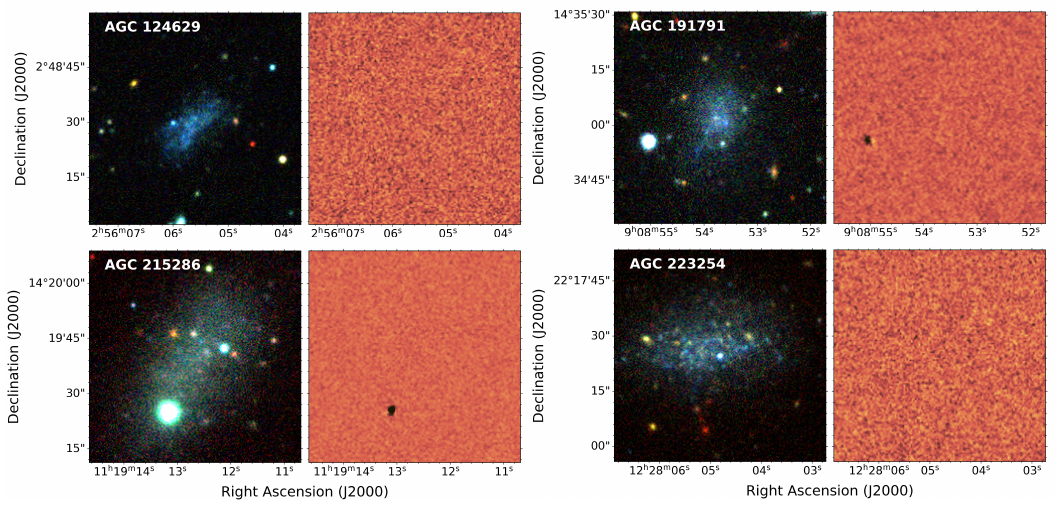}
\caption{Images of four representative SHIELD galaxies that were not detected in our narrowband H$\alpha$ images.  For each galaxy we show a pair of images, consisting of a color composite image created using {\it giz} DESI Legacy imaging data (left) and the continuum-subtracted H$\alpha$ narrowband image (right).  Each image segment covers an area of roughly 56 $\times$ 56 arcsec, and in all cases N is up and E is left.}
\label{fig:SHIELDimages7}
\end{figure*}

\begin{figure*}
\centering
\includegraphics[width=7.1in]{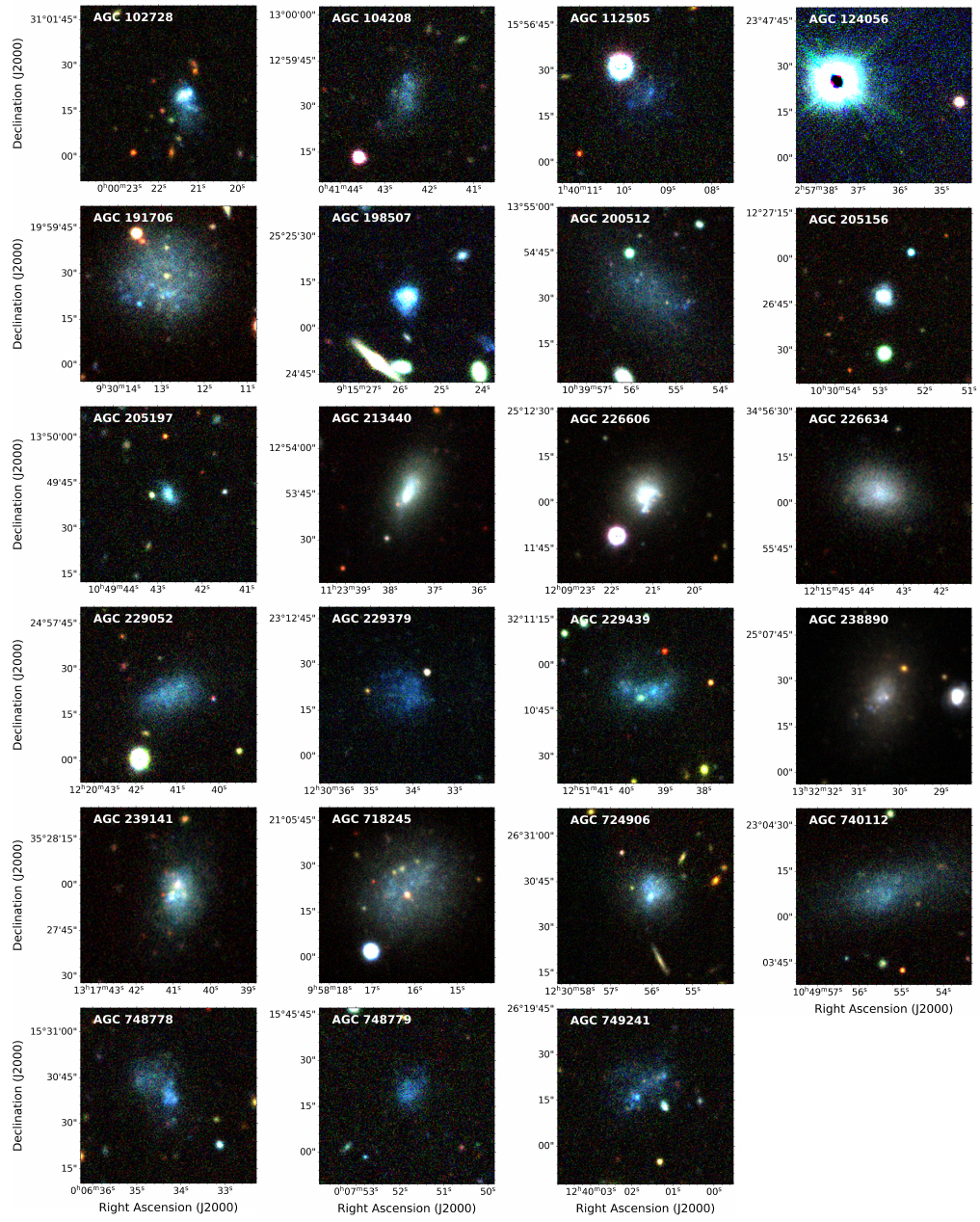}
\caption{ Images of the remaining 23 SHIELD galaxies that were not detected in our narrowband H$\alpha$ images.  For each galaxy we show only the color composite image created using {\it giz} DESI Legacy imaging data.  Each image segment covers an area of roughly 56 $\times$ 56 arcsec, and in all cases N is up and E is left.  The extremely low surface brightness system AGC 124056 (top row, rightmost image) is barely detectable in our images due the bright foreground star located in the field.  }
\label{fig:SHIELDimages8}
\end{figure*}

Since this paper is one of the first to present an analysis of the full sample of SHIELD dwarfs, it seems appropriate to exhibit images for all 82 SHIELD galaxies.  The imaging data for the SHIELD galaxies are shown in Figures~\ref{fig:SHIELDimages1} -~\ref{fig:SHIELDimages8}.  

Images of the 55 H$\alpha$-detected SHIELD galaxies are presented (Figures~\ref{fig:SHIELDimages1} -~\ref{fig:SHIELDimages6}). Both broadband color composite images created using the DESI legacy {\it giz} images \citep{dey_2019} and our continuum-subtracted H$\alpha$ images are shown.  The remaining 27 galaxies are the H$\alpha$ non-detections.  Four of these are shown using the same format as the H$\alpha$ detections (both broadband and continuum-subtracted narrowband images) in order to give the reader a sense of the depth of the H$\alpha$ non-detections (Figure~\ref{fig:SHIELDimages7}).  For the remaining 23 H$\alpha$ non-detections we present only the DESI {\it giz} color composite broadband images (Figure~\ref{fig:SHIELDimages8}).

It is not our intention to evaluate the morphology of each SHIELD galaxy individually in the current paper.  Rather, our goal in presenting these images is to provide the reader with a sense of the range of the optical characteristics of the galaxies that make up this unique \ion{H}{1}-selected sample.  To that end, we limit our discussion in the remainder of this section to highlighting the contrasting optical and narrowband morphologies of the SHIELD galaxies.

First, we note that the broadband images reveal a wide range of surface brightness (SB) levels for the stellar components of the SHIELD galaxies.  We assigned each SHIELD galaxy into one of three rough SB categories of high (e.g., AGC 194019 and AGC 198691 in Figure~\ref{fig:SHIELDimages2}), medium (e.g., AGC 103722 and AGC 110482 in Figure~\ref{fig:SHIELDimages1}), and low SB (e.g., AGC 112521 and AGC 123352 in Figure~\ref{fig:SHIELDimages1}), and found that there are approximately equal numbers in each category among the 55 H$\alpha$-detected galaxies.  We stress that these SB categories are not defined quantitatively, and are meant only to provide a broad classification.  

We find the extreme diversity in the optical characteristics of the galaxies to be quite interesting, and connect it directly to the \ion{H}{1} selection of the original ALFALFA sample.  The optical surface brightness of any given SHIELD galaxy appears to be largely independent of the presence of available gas.  This point has been noted in previous studies, such as \citet{vanzee1996, vanzee1997}, \citet{vanzee2001}, and \citet{cannon2009}.  The reservoirs of cold \ion{H}{1} gas present in the SHIELD dwarfs do not necessarily translate into a uniformly high star-formation efficiency that would in turn result in intermediate and high stellar surface brightnesses in all cases.

Another aspect of the \ion{H}{1} selection of the SHIELD sample is the ability to include galaxies that might otherwise be missed in optical searches.  This naturally includes the extremely low SB dwarfs such as AGC 123352 in Figure~\ref{fig:SHIELDimages1} (right column, 4th from top) and AGC 229053 in Figure~\ref{fig:SHIELDimages4} (right column, 3rd from top).  It also includes a handful of galaxies that are largely hidden behind the glare of foreground stars: AGC 124635 in Figure~\ref{fig:SHIELDimages1} (left column, bottom row), 
AGC 229090 in Figure~\ref{fig:SHIELDimages4} (left column, 4th from top), and, most notably,
AGC 124056 in Figure~\ref{fig:SHIELDimages8} (rightmost column, top row).  This latter galaxy would almost certainly never have been detected in optical surveys, since it is nearly invisible even in our pointed observations.

While the SBs exhibited by the H$\alpha$-detected SHIELD galaxies are fairly evenly distributed between the three levels of our simple classification scheme, this is not the case for the H$\alpha$ non-detections (Figures~\ref{fig:SHIELDimages7} and~\ref{fig:SHIELDimages8}).  Here we find that a majority (18 of 27) are assigned to the low SB group.  This is no surprise, given that low surface brightness (LSB) galaxies tend toward having lower star-formation activity.  There are, however, a number of higher SB systems among the H$\alpha$ non-detections, such as AGC 205156 (Figure~\ref{fig:SHIELDimages8}, right column, second row), AGC 213440, and AGC 226606 (Figure~\ref{fig:SHIELDimages8}, middle two columns, third row).  All three of these galaxies were clearly detected in the FUV \citep{gormanous26}, indicating significant star formation within the past $\sim$100 Myr.  However, we detect no evidence of recent star formation via our H$\alpha$ images. 

Next we draw the reader's attention to the H$\alpha$ morphologies of the SHIELD galaxies.  Since the sample consists exclusively of low-mass galaxies, it is not a surprise to find that the overall appearances of the galaxies in the light of their ionized gas tends to be rather simple.  A census of the full sample reveals that 30 of the 55 H$\alpha$-detected SHIELD galaxies possess a single \ion{H}{2} region / star-forming complex, while another 10 galaxies exhibit two star-forming regions.  Examples of SHIELD galaxies that possess a single \ion{H}{2} region / star-forming complex include AGC 103722, AGC 111164, and AGC 112521 (all shown in Figure~\ref{fig:SHIELDimages1}).  

SHIELD galaxies that have two \ion{H}{2} regions / star-forming complexes include AGC 124635 (Figure~\ref{fig:SHIELDimages1}) and AGC 198712 (Figure~\ref{fig:SHIELDimages2}).  AGC 174605 (Figure~\ref{fig:SHIELDimages2}) possesses two rather faint \ion{H}{2} regions plus some additional diffuse emission. We identify eight SHIELD galaxies as harboring three or more regions of current star formation.  Examples of galaxies that show these more complex H$\alpha$ morphologies include AGC 731457 (Figure~\ref{fig:SHIELDimages5}) and AGC 749237 (Figure~\ref{fig:SHIELDimages6}).  

While in many cases the \ion{H}{2} regions observed in the SHIELD galaxies are located at or near their optical centers, there are several systems that possess bright knots of H$\alpha$ emission that are located near the outer edges of the stellar disks.  For example, Figure~\ref{fig:SHIELDimages3} includes two cases of ``edge" \ion{H}{2} regions: AGC 210960 and AGC 215284 exhibit bright blue knots near the outer edges of their broadband images, both of which coincide with strong, compact regions of H$\alpha$ emission.  Three additional galaxies with similar morphologies are seen in Figure~\ref{fig:SHIELDimages4}: AGC 223231, AGC 224312, and AGC 239031.

Many systems show diffuse H$\alpha$ emission, sometimes in conjunction with discrete \ion{H}{2} regions and sometimes as the only source of nebular emission. Galaxies with diffuse emission are fairly common within the SHIELD sample.  For example, in Figure~\ref{fig:SHIELDimages1} we see several galaxies that exhibit diffuse emission while showing a range of H$\alpha$ morphologies: AGC 110482 possesses 5 discrete \ion{H}{2} regions as well as three areas of diffuse emission.  AGC 111164 exhibits some diffuse H$\alpha$ emission just to the right (west) of the single bright \ion{H}{2} region.  AGC 111977 has a single faint \ion{H}{2} region near the southern end of the stellar distribution, with an extensive complex of diffuse H$\alpha$ filaments extending further to the southwest.  AGC 123352 is an extreme LSB system with a single ultra-faint \ion{H}{2} region surrounded by additional diffuse emission.  In sharp contrast, AGC 171459 possesses a high SB stellar component but only a modest amount of diffuse emission coincident with the centroid of the stellar emission.

The origin of the diffuse H$\alpha$ emission is unclear.  Two commonly suggested physical processes that have been invoked in the past include shocks and leakage of Lyman continuum photons from \ion{H}{2} regions into the surrounding low density gas.  In spiral galaxies most diffuse ionized gas is typically associated with star-formation in or near the spiral arms \citep[e.g.,][]{walterbos1994, mihos2024}.  The situation for low-mass dwarf galaxies is less clear.  In the case of the SHIELD galaxies, our interpretation tends to lean toward the shock hypothesis, for a number of reasons.  First, in several cases SHIELD galaxies possess extended diffuse emission in galaxies lacking any bright, compact \ion{H}{2} regions.  Examples include AGC 171459 (Figure~\ref{fig:SHIELDimages1}) and AGC 728909 (Figure~\ref{fig:SHIELDimages5}).  These galaxies have no apparent source for the origin of any escaping Lyman continuum photons.  In other cases, the diffuse emission is not located in close proximity to any sources of ionizing photons (e.g., AGC 110482 in Figure~\ref{fig:SHIELDimages1}).  Third, in the nearby low-mass gas-rich galaxy Leo P \citep{rhode2013,mcquinn2015c}, which has properties similar to many of the SHIELD galaxies, deep VLT/MUSE data presented in \citet{evans2019} clearly show that the diffuse emission is due to SN shocks.  While none of these points rule out the possibility that the diffuse emission may be due at least in part to escaping Lyman continuum radiation in {\it some} of the SHIELD galaxies, it suggests that shocks are the more likely excitation source in many cases.

We stress that the lack of depth and poor spatial resolution of our H$\alpha$ images makes a definitive interpretation regarding the nature of the diffuse emission difficult in many cases.  It is entirely possible that our current set of images are missing substantial amounts of unseen diffuse emission.  A more comprehensive analysis on this topic will require deeper images obtained with a larger aperture telescope.

We also point out that several of the SHIELD galaxies are barely detected in our NB images.  Examples include AGC 205590 (one very faint \ion{H}{2} knot), AGC 208477 (one very faint \ion{H}{2} knot), and AGC 210220 (weak diffuse emission), all seen in Figure~\ref{fig:SHIELDimages3}.  AGC 229053 (Figure~\ref{fig:SHIELDimages4}) is an extreme LSB system with four or five very faint \ion{H}{2} regions.  Perhaps the most elusive of all of the putative H$\alpha$ detections is AGC 728909 (Figure~\ref{fig:SHIELDimages5}) which shows faint diffuse emission centered on the relatively high SB stellar component.

One of the best known and most studied of the SHIELD galaxies is AGC 198691 (a.k.a.\ the Leoncino dwarf, named for its location in the constellation Leo Minor).  It was recognized shortly after its discovery as being one of the lowest metallicity galaxies known \citep{hirschauer2016}, with an oxygen abundance of only $\sim$2\% solar \citep[see][for more recent metallicity studies of Leoncino]{aver2022,aver2026,rogers2026}.  It is shown in Figure~\ref{fig:SHIELDimages2} as an extremely compact system with a diameter of $\sim$0.8 kpc \citep{mcquinn2020} and a single bright knot of H$\alpha$ emission.

A key motivation for acquiring the H$\alpha$ images was to provide targets for subsequent spectroscopic observations with the goal of measuring nebular abundances in sources with sufficiently strong emission.  Future papers by this group will present these spectra together with derived metallicity estimates for many of the SHIELD galaxies.

\subsection{H$\alpha$ SFR Properties}\label{sec:HAsfr}

\begin{figure*}
\centering
\includegraphics[width=6.0in]{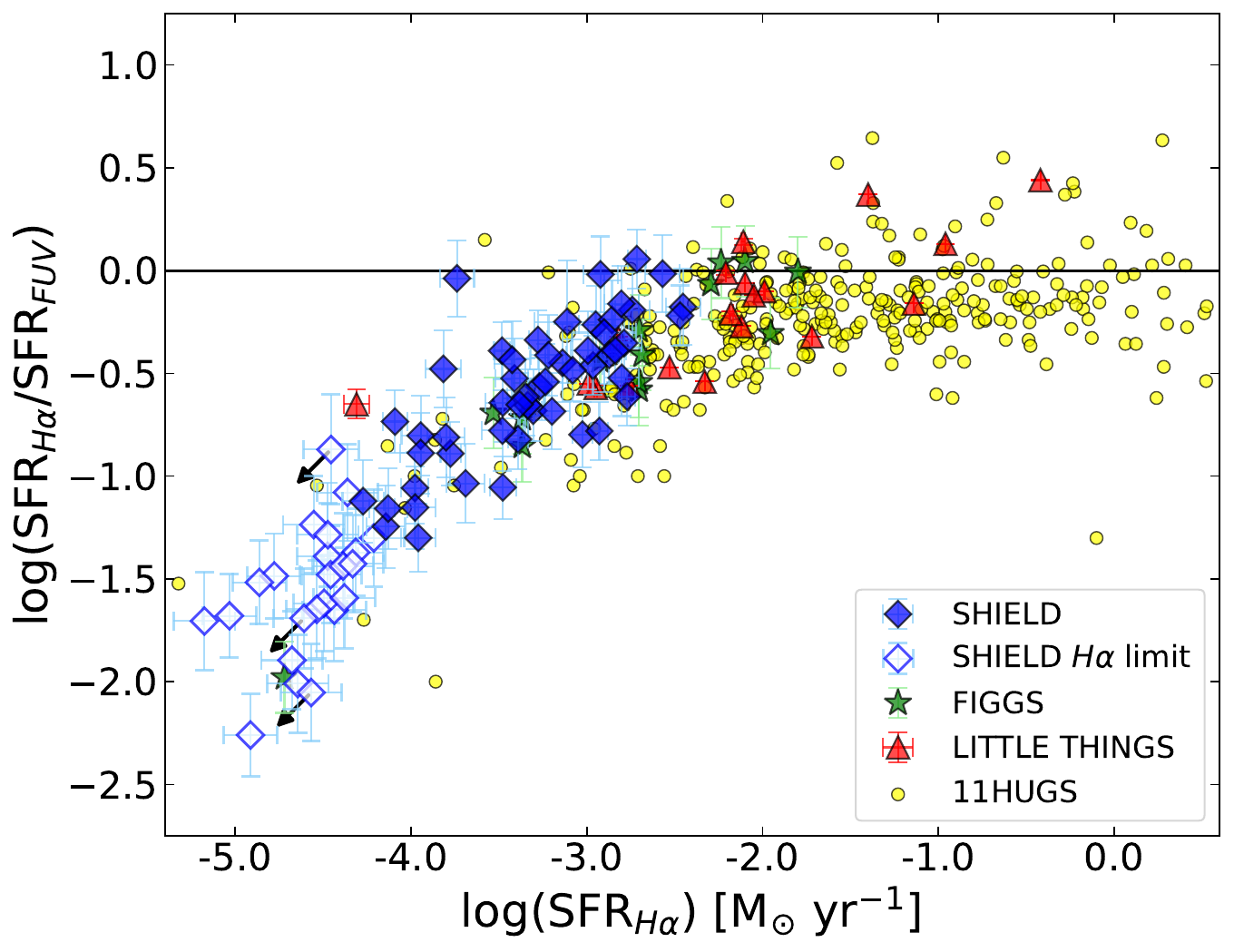}
\caption{ Log-log graph of (H$\alpha$ SFR/FUV SFR) vs.\ H$\alpha$ SFR, in units of solar masses per year, for SHIELD as well as other nearby galaxy surveys. The open diamonds indicate SHIELD galaxies whose H$\alpha$ fluxes represent 3$\sigma$ upper limits; upper-limit arrows are shown on only a few of these open diamonds. The solid line represents equality (i.e. H$\alpha$ SFR = FUV SFR).  Above log(SFR$_{H\alpha}$) $\sim$ $-$2.5 there is a good agreement between the two SFRs, while below this value the instantaneous SFR (SFR$_{H\alpha}$) becomes an increasingly poor indicator of the time-averaged recent SFR (SFR$_{FUV}$).}
\label{fig:SFRcomparisonplot1}
\end{figure*}

In this subsection we explore a number of relationships between the newly derived H$\alpha$ SFRs for the SHIELD galaxies and several relevant physical parameters.  As part of this analysis we directly compare the properties of the SHIELD dwarfs with those of three representative comparison samples of other galaxies in the nearby universe.  Much of what is presented here follows the analysis in \citet{gormanous26}, but using H$\alpha$ SFRs rather than FUV SFRs.  

The selection of our particular comparison samples exactly mimics the choices made in \citet{teich2016}, which were then carried over to \citet{gormanous26}.   We do this to provide continuity between the previous and current SHIELD-based studies.  Our comparison data sets are drawn from: 11HUGS \citep{Lee_2009b}, LITTLE THINGS \citep{hunter2012}, and FIGGS \citep{Roychowdhury_2014}.  
More details about the comparison samples can be found in \citet{gormanous26}.  Here we simply note that the LITTLE THINGS and FIGGS catalogs consist primarily of nearby dwarf galaxies, while 11HUGS serves as a comprehensive optical-flux limited sample of galaxies in the local universe out to distances of 11 Mpc.  We note that we do not utilize the VLA-ANGST \citep{ott2012} dwarf galaxy sample in the current paper.  While VLA-ANGST was an important comparison data set used in both \citet{teich2016} and \citet{gormanous26}, this sample lacks H$\alpha$ measurements which makes it inappropriate for use in the current study.

All plots presented in this subsection and the next (Section~\ref{sec:HAsSFR}) utilize the same plotting symbols to designate galaxies from the various samples: 11HUGS are yellow circles, LITTLE THINGS are red triangles, FIGGS are green five-pointed stars, and SHIELD are blue diamonds.  For SHIELD galaxies that are detected in H$\alpha$, the diamonds are filled, while the H$\alpha$ non-detections are unfilled.

\subsubsection{H$\alpha$ and FUV SFR Ratio Comparisons}\label{sec:sfrratio}

We begin by comparing the logarithm of the ratio of H$\alpha$ to FUV SFR vs.\ log(SFR$_{H\alpha}$) in Figure~\ref{fig:SFRcomparisonplot1}. The solid black line represents equality (i.e., H$\alpha$ SFR = FUV SFR) and the black arrows represent the direction of the upper-limit values.  For the sake of clarity, these upper-limit arrows are placed on only a few of the open diamonds. The plot shows the familiar result that the H$\alpha$ SFR starts to underestimate the actual SFR as lower SFR values are reached \citep[e.g.,][]{Lee_2009b, Hunter_2010, fumagalli2011, weisz2012, Roychowdhury_2014, gormanous26}.  This plot is similar to Figure 5 of \citet{gormanous26}, which compares the logarithm of the ratio of H$\alpha$ to FUV SFR vs.\ log(SFR$_{FUV}$). The SHIELD galaxies densely populate the region below log(SFR$_{H\alpha}$) $\sim$ $-$2.5 in the figure, providing improved definition of the divergence between the two SFR indicators.

The departure from a flat trend in SFR$_{H\alpha}$/SFR$_{FUV}$ appears to begin near log(SFR$_{H\alpha}$) = $-$2.5. Above this departure value, the distribution of log(SFR$_{H\alpha}$/SFR$_{FUV}$) is rather flat, with most galaxies being located between 0.0 and $-$0.5, but with substantial scatter above and below these values.  At log(SFR$_{H\alpha}$) $\approx$ $-$2.5 the downward trend in the SFR ratio starts to become evident, while by log(SFR$_{H\alpha}$) = $-$3.0 it appears to be quite clear.  The divergence between SFR$_{H\alpha}$ and SFR$_{FUV}$ continues to increase as lower SFRs are reached.  By log(SFR$_{H\alpha}$) = $-$5.0 the characteristic difference between SFR$_{H\alpha}$ and SFR$_{FUV}$ reaches a factor of $\sim$100.  Potential causes of this deviation that have been suggested include stochastic sampling of the high-mass end of the stellar initial mass function \citep{fumagalli2011, dasilva2012}, bursty star-formation histories \citep{Lee_2009b,weisz2012}, the leakage of ionizing photons from star-forming regions \citep{oey2007} and a non-constant IMF \citep{meurer2009, Gunawardhana2011}.

\begin{figure}
\centering
\includegraphics[width=3.35in]{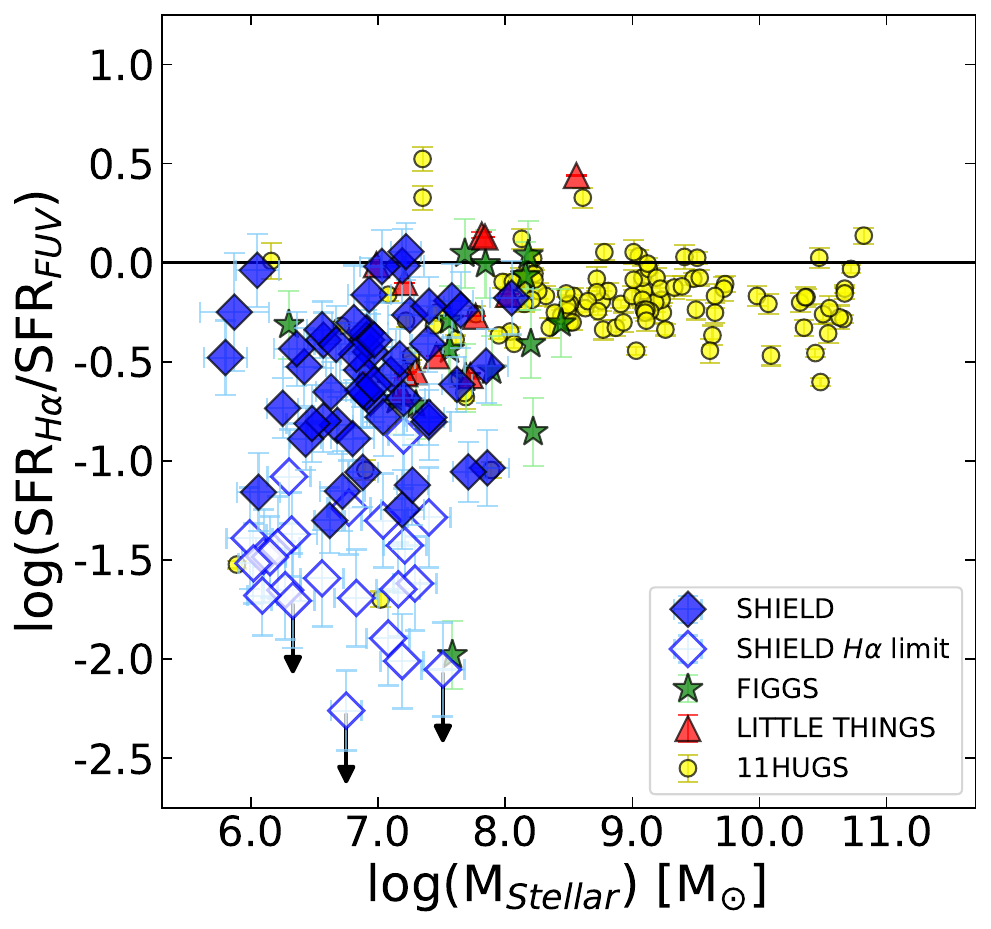}
\caption{Log-log graph of (H$\alpha$ SFR/FUV SFR) vs.\ Stellar Mass. The solid line represents equality.  For stellar masses above $\sim$10$^8$ M$_\odot$, the two SFR indicators agree reasonably well with each other.  At masses below this value, the H$\alpha$ SFR/FUV SFR ratio can exhibit a wide range of values between 1.00 and 0.01, but with no obvious correlation with M$_{stellar}$.}
\label{fig:ratiovsstellarmass}
\end{figure}

\begin{figure*}
\centering
\includegraphics[width=6.60in]{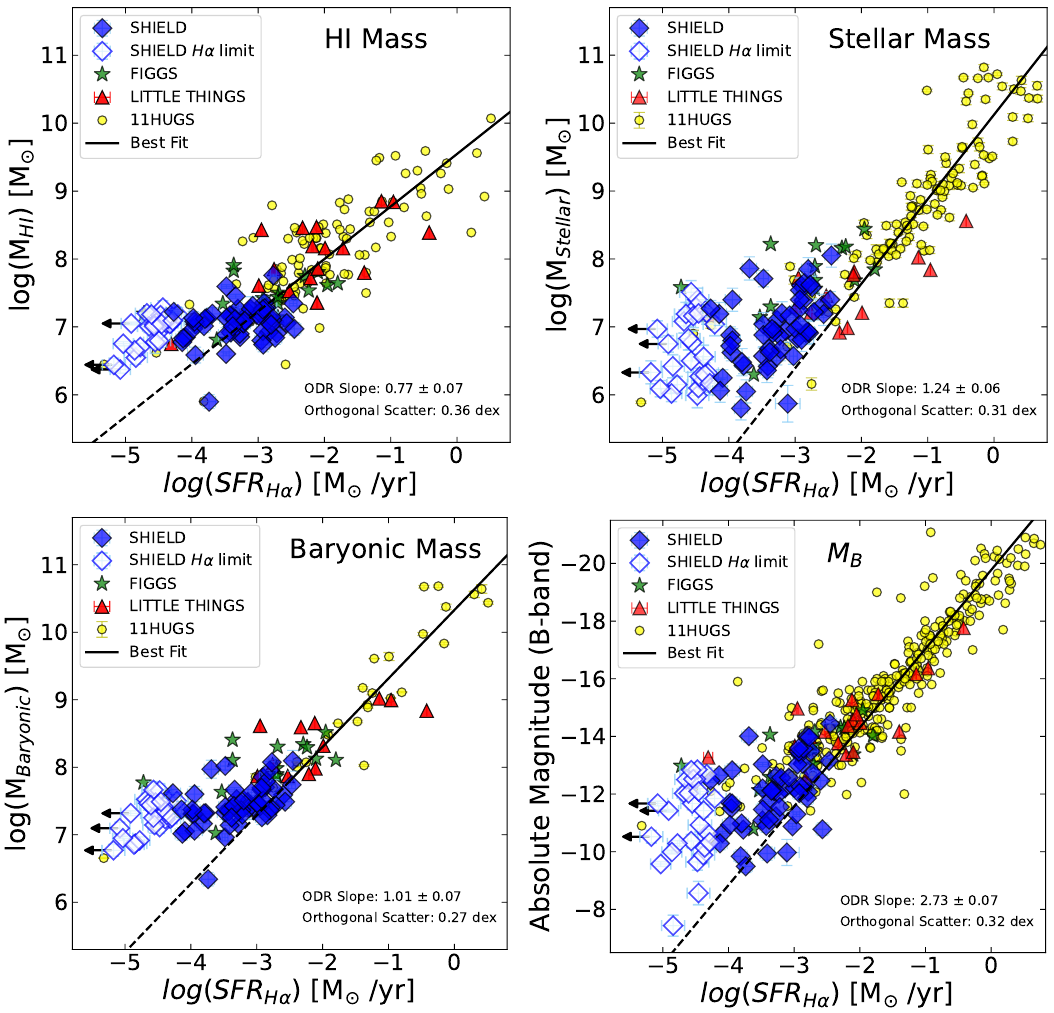}
\caption{Log-log plots of mass and absolute magnitude vs.\ H$\alpha$ SFR for the SHIELD galaxies as well as our comparison sample galaxies.  {\it upper left}: \ion{H}{1} mass vs.\ H$\alpha$ SFR; {\it upper right}: stellar mass vs.\ H$\alpha$ SFR; {\it lower left}: baryonic mass vs.\ H$\alpha$ SFR; {\it lower right}: B-band absolute magnitude vs.\ H$\alpha$ SFR.   In each plot, the solid line indicates an orthogonal distance regression best fit for the data with log(SFR$_{H\alpha}$) $> -2.5$, the approximate value at which the FUV and H$\alpha$ SFRs start to diverge significantly (e.g., Figure~\ref{fig:SFRcomparisonplot1}).  The slopes and scatters about the best-fit lines are indicated.  The dashed lines are extensions of these best fit lines, and are meant solely to guide the eye.  In all four plots, one sees a reasonably strong correlation for values of  log(SFR$_{H\alpha}$) above $-$2.5, but the correlations break down at lower SFRs. }
\label{fig:massvsSFR}
\end{figure*}

Many authors have noted that the divergence between SFR$_{H\alpha}$ and SFR$_{FUV}$ becomes more pronounced in low-mass galaxies, where effects like bursty star-formation histories and stochastic IMF sampling are expected to be have their largest impacts \citep[e.g.,][]{Lee_2009b, fumagalli2011, weisz2012}.  However, this does not necessarily mean that the SFR$_{H\alpha}$/SFR$_{FUV}$ divergence is directly correlated with galaxy mass.  To explore this issue, we examine the logarithm of the H$\alpha$-to-FUV SFR ratio as a function of stellar mass in Figure~\ref{fig:ratiovsstellarmass}.  At masses of M$_{Stellar}$ $\gtrsim$ 10$^{8.0}$ M$_\odot$ values of the SFR$_{H\alpha}$/SFR$_{FUV}$ ratio exhibit a flat distribution, with most galaxies lying  between log(SFR$_{H\alpha}$/SFR$_{FUV}$) of 0.0 and $-$0.5.  This level of scatter in the SFR ratio at higher masses has been noted previously (e.g., see Figure 3 of \citet{weisz2012}).

However, for M$_{Stellar}$ $\lesssim$ 10$^{8.0}$ M$_\odot$, the appearance of the plot changes dramatically.  Here, values of log(SFR$_{H\alpha}$/SFR$_{FUV}$) cover the full range from 0.5 to $-$2.3.  Furthermore, log(SFR$_{H\alpha}$/SFR$_{FUV}$) is no longer correlated with stellar mass.  Interestingly, some of the lowest mass SHIELD galaxies have values of the H$\alpha$-to-FUV SFR ratio that are similar to the highest mass 11HUGS galaxies.  Again, this same result has been noted previously by \citet{weisz2012}.

Figure~\ref{fig:ratiovsstellarmass} strongly suggests that the deviation between H$\alpha$ and FUV SFRs is not caused by the mass of the galaxy alone, but rather is moderated by the rate of star formation.  Low mass galaxies with high current SFRs have H$\alpha$ and FUV SFRs that are in reasonable agreement with each other, while low mass galaxies with low current SFRs will show strong divergence between the two SFR indicators.  Naturally, lower mass galaxies tend to have lower SFRs, so it has become common practice to associate low mass galaxies with large differences in their H$\alpha$ and FUV SFRs.  However, for star-forming dwarfs that undergo bursts or cycles of vigorous star formation {\it and} are observed during a ``high state" in this cycle (e.g., blue compact dwarfs (BCDs)), H$\alpha$ measurements can provide a fairly representative estimate of the time-averaged recent SFR (i.e., within a factor of $\sim$two).

It is not the goal of the current paper to carry out a detailed critique of the possible reasons for the large disagreements between SFR(H$\alpha$) and SFR(FUV) as illsutrated in Figure~\ref{fig:SFRcomparisonplot1}.  However, the result shown in Figure~\ref{fig:ratiovsstellarmass} -- that the deviation between H$\alpha$ and FUV SFRs correlates more strongly with SFR than with galaxy mass -- would seem to be a strong argument in favor of the picture that the differences between the two SFR measurements is caused primarily by the stochastic time variability in the star-formation in these low-mass systems \citep[e.g.,][]{Lee_2009b, weisz2012}.   This does not rule out the possibility that some of the other factors mentioned above may also contribute to the amplitude of the deviations, but we suggest that at best these other factors are relatively minor contributors to the observed deviations.

\subsubsection{Mass Quantities vs. H$\alpha$ SFR}\label{sec:massplots}

\citet{gormanous26} carried out a linear regression analysis that used the SHIELD and comparison galaxy samples to examine the correlations between FUV SFR and four different mass tracers: \ion{H}{1} mass, stellar mass, baryonic mass, and $B$-band absolute magnitude. They considered the absolute magnitude to be a mass-like quantity, as it is often seen as an analog to the stellar mass of a galaxy.  They found that SFR$_{FUV}$ is strongly correlated with each of these tracers over the full mass, luminosity and SFR ranges present in the SHIELD plus comparison samples.  Furthermore, they found that the tightest relation between mass and FUV SFR occurs for the baryonic mass.

Here we carry out a similar analysis using H$\alpha$ SFR.  Figure~\ref{fig:massvsSFR} shows log–log comparisons of \ion{H}{1} mass, stellar mass, baryonic mass, and $B$-band absolute magnitude versus H$\alpha$ SFR. In each panel of Figure~\ref{fig:massvsSFR}, the solid line represents the best-fit orthogonal distance regression (ODR) relation for galaxies with log(SFR$_{H\alpha}$) $> -2.5$, the approximate value below which we determined that the FUV and H$\alpha$ SFRs start to diverge significantly (Section~\ref{sec:sfrratio}). The dashed lines are extensions of these fits for clarity. The corresponding slopes, uncertainties, and orthogonal scatters are indicated in each panel.

Across all four relations, the data follow an approximately linear trend at higher SFRs, with a clear deviation emerging below log(SFR$_{H\alpha}$) $\sim -2.5$. This low-SFR regime is densely populated by SHIELD galaxies, including all sources with H$\alpha$ upper limits.  It is important to note that the linear fits only include the galaxies in the higher mass, higher SFR portion of the plots and exclude most of the SHIELD galaxies and the dwarfs from the comparison samples.   Hence, the results of our fitting procedure should not be compared with the fits from \citet{gormanous26}.   Our main purpose for including these fits in Figure~\ref{fig:massvsSFR} is to establish a trend line for galaxies with relatively high H$\alpha$ SFRs against which the departure from linearity at lower SFRs can clearly be seen.

The results displayed in all four panels of Figure~\ref{fig:massvsSFR} are fully consistent with our findings from Section~\ref{sec:sfrratio}.  At log(SFR$_{H\alpha}$) below $-$2.5, the SHIELD and comparison galaxies scatter to the left of the trend lines, indicating that the instantaneous H$\alpha$ SFR is increasingly underestimating the time-averaged recent SFR.  However, the deviation between the H$\alpha$ SFR from the fiducial fit lines is not directly related to stellar mass.  Rather, some of the SHIELD and dwarf comparison galaxies follow the linear regression lines down to very low masses and faint absolute magnitudes, consistent with the results seen in Figure~\ref{fig:ratiovsstellarmass}.   This is perhaps easiest to visualize in the baryonic mass vs. H$\alpha$ SFR plot in Figure~\ref{fig:massvsSFR} (lower left panel), where the galaxy with the lowest value of M$_{baryonic}$ (AGC 732009) falls close to the extension of the linear fit.

\begin{figure}
\centering
\includegraphics[width=3.35in]{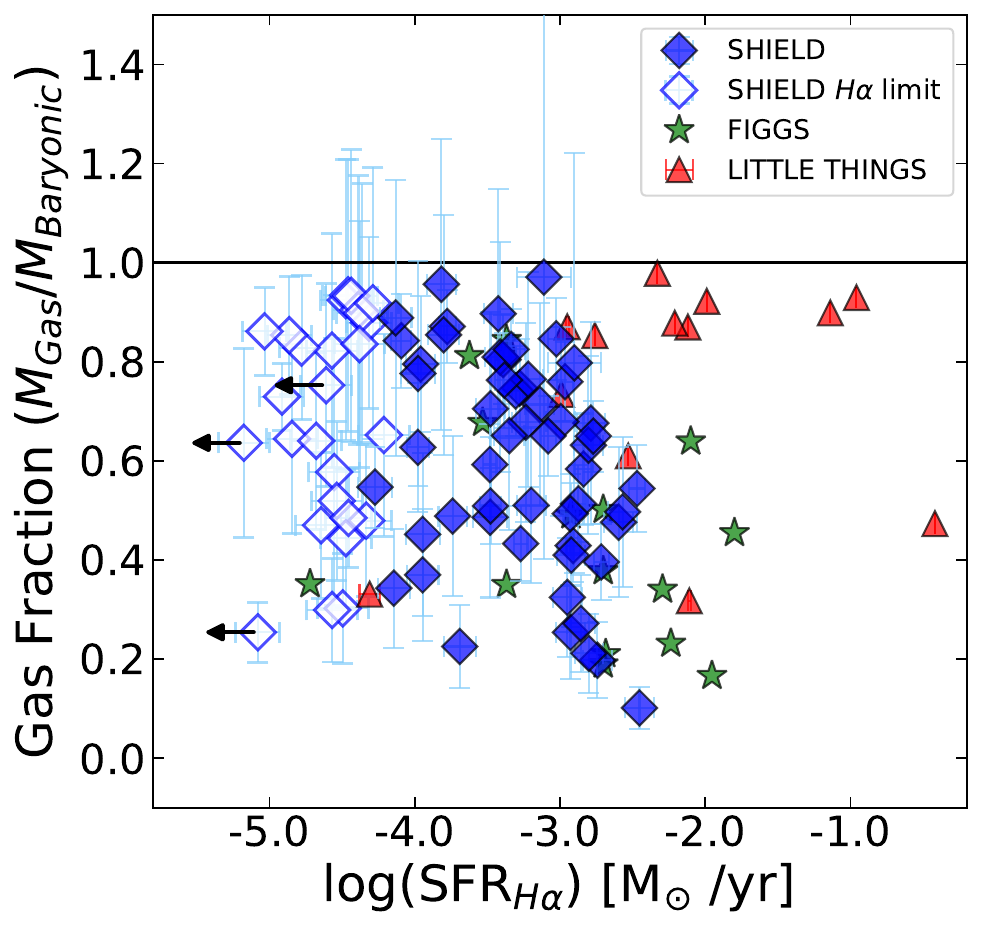}
\caption{Graph of baryonic gas mass fraction (M$_{gas}$ / M$_{baryonic}$) vs. the logarithm of the H$\alpha$ SFR.  The horizontal solid line indicates M$_{gas}$ = M$_{baryonic}$, which would be the case for a starless galaxy.  There is no correlation between the amount of gas available for making stars and the H$\alpha$ SFR.}
\label{fig:BaryonicFractionVssfr}
\end{figure}

\begin{figure*}
\centering
\plottwo{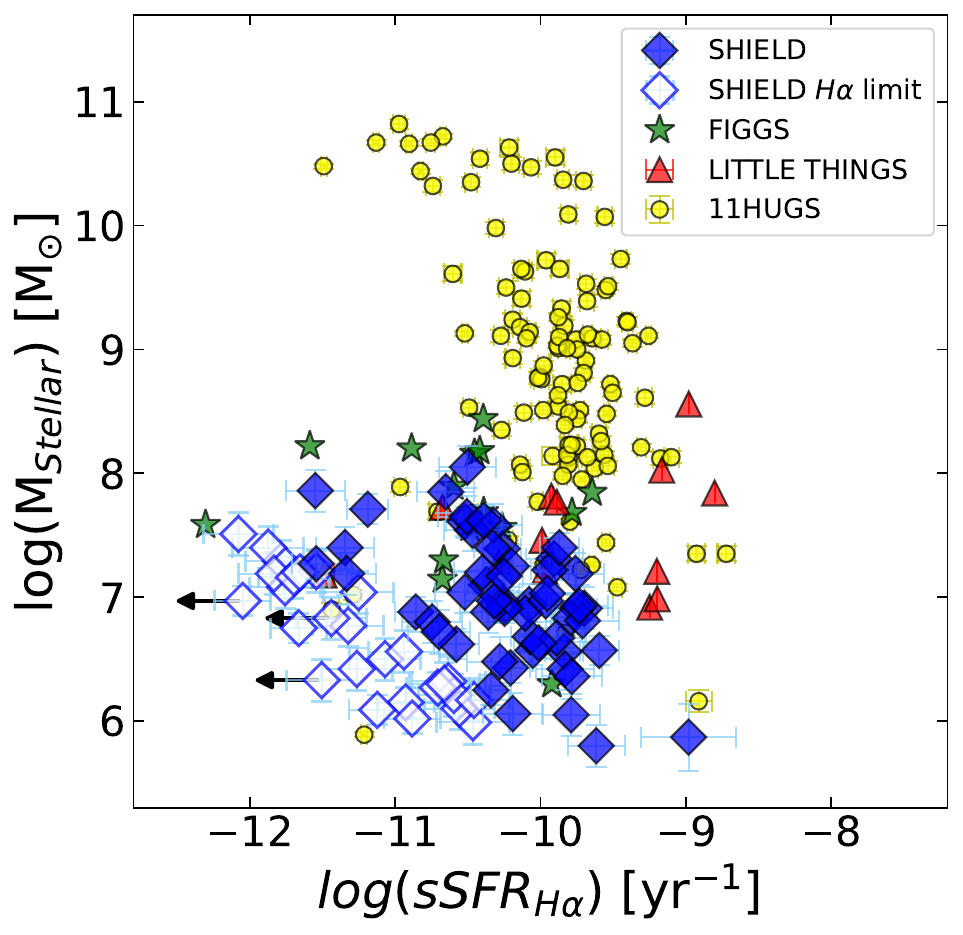}{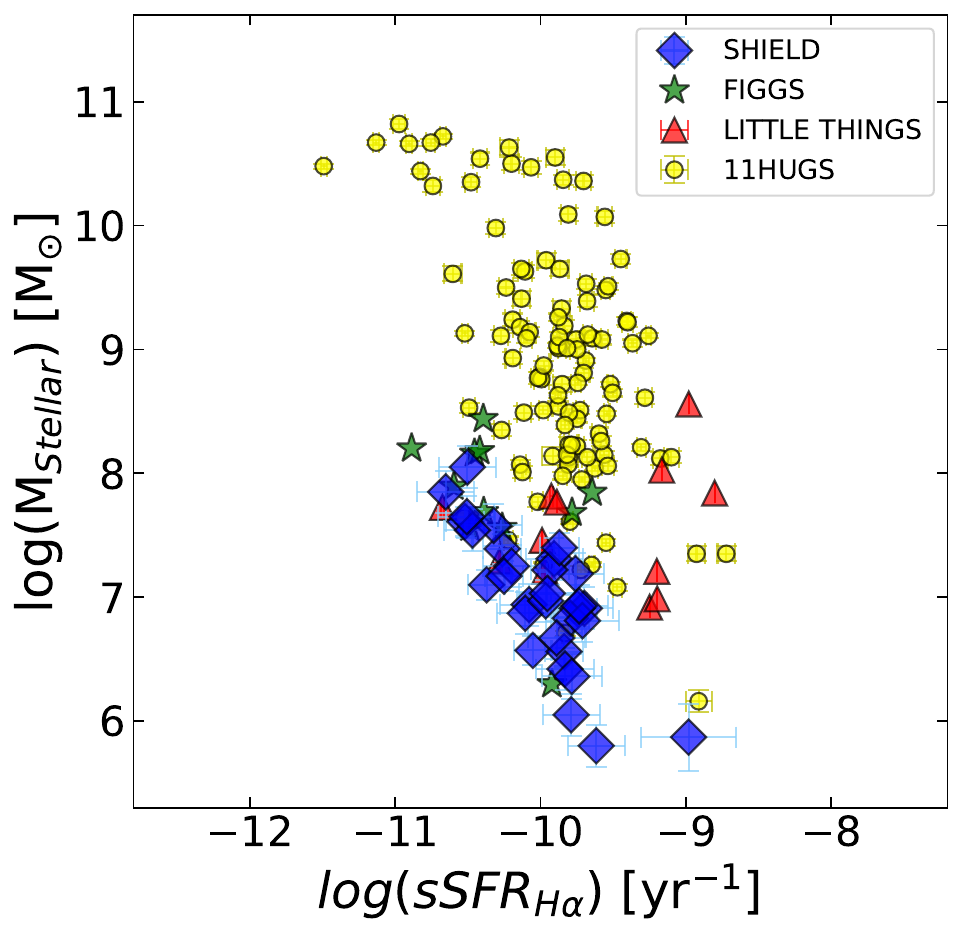}
\caption{Log-log plots of Stellar Mass vs.\ H$\alpha$ sSFR.  ({\it Left}) The full sample of SHIELD galaxies, including H$\alpha$ upper limits, as well as our comparison samples used in previous plots.  ({\it Right}) The same quantities and samples plotted, but with galaxies with log(SFR$_{H\alpha}$/SFR$_{FUV}$) $<$ -0.6 removed.  This cut removes those galaxies with values of SFR$_{H\alpha}$ that are not in reasonable agreement with their time-averaged recent SFRs.  See text for details.  The SHIELD galaxies in the culled sample are seen to exhibit sSFR$_{H\alpha}$ values that overlap those of the galaxies in the local comparision samples.  Not surprisingly, there are no extreme star-forming galaxies (i.e., log(sSFR$_{H\alpha}$) $\sim$ $-$8 yr$^{-1}$ or higher) in this group of nearby galaxies.}
\label{fig:StellarMassVsssfr2}
\end{figure*}

\subsubsection{Baryonic Gas Mass Fraction vs. H$\alpha$ SFR}\label{sec:gasfractionvsHAsfr}

Our next comparison is between the baryonic gas mass fraction $(M_{Gas} / M_{baryonic})$ and the H$\alpha$ SFR, which is shown in Figure~\ref{fig:BaryonicFractionVssfr}.  Here we include only the comparison galaxies from the two dwarf-dominated samples: LITTLE THINGS and FIGGS.  In contrast to the mass-related quantities examined previously, the gas fraction exhibits no strong correlation with H$\alpha$ SFR. Galaxies with detected H$\alpha$ fluxes span more than three orders of magnitude of H$\alpha$ SFR at any given gas fraction, and over four orders of magnitude when including H$\alpha$ upper limit fluxes.  Likewise, at any given H$\alpha$ SFR (except the very highest) the galaxies shown in Figure~\ref{fig:BaryonicFractionVssfr} uniformly cover the gas mass fractions from 0.2 to 1.0, indicating that gas fraction alone is not a useful predictor of recent star formation activity. The SHIELD galaxies extend the relation to lower SFRs than the comparison samples, but do not reveal a distinct turnoff analogous to that observed in the stellar, \ion{H}{1}, and baryonic mass relations. 

This behavior is consistent with the results of Figure 11 in \citet{gormanous26}, where we found no clear relationship between baryonic gas mass fraction and FUV SFR. Taken together, these results indicate that the amount of gas present relative to a galaxy's total baryonic mass is not a primary driver of either recent or ongoing star formation activity. The broader range of H$\alpha$ SFRs observed here, coupled with the gas-dominated nature of the SHIELD galaxies \citep{gormanous26}, further suggests that the regulation of star formation in these low-mass systems depends on factors beyond the overall gas supply alone.  One possible explanation for this result might be the suggestion that a galaxy's angular momentum content can affect the conversion of gas into stars \citep[e.g.,][]{obreschkow2016, lutz2018, mancerapina2021b}.  The gas disk of a galaxy with high specific angular momentum will be more stable against gravitational collapse, allowing for higher gas mass fractions while keeping the SFRs at low to moderate levels.

\subsection{H$\alpha$ Specific Star-Formation Rate (sSFR) Properties }\label{sec:HAsSFR}

Following \citet{gormanous26}, we next present plots that illustrate the specific star-formation rates (sSFRs) of the SHIELD galaxies as well as the galaxies in our comparison sample.  Here we focus on the H$\alpha$ sSFRs, despite the fact that for many of the lower-mass dwarfs the H$\alpha$ SFRs are underestimating the time-averaged recent SFRs by substantial amounts (e.g., Figures~\ref{fig:SFRcomparisonplot1} and~\ref{fig:ratiovsstellarmass}).

Figure~\ref{fig:StellarMassVsssfr2} plots the logarithm of the stellar mass vs. the logarithm of sSFR$_{H\alpha}$ for the SHIELD dwarfs as well our comparison samples.  To clearly illustrate the impact caused by the H$\alpha$ SFRs we present two versions of the plot.  First, we show on the left the entire set of SHIELD galaxies plus all comparison galaxies with the necessary data to be included in the plot.  This includes all the SHIELD galaxies with upper limits for SFR$_{H\alpha}$.  A large fraction of the dwarf galaxies exhibit extremely low sSFR$_{H\alpha}$ values ($<$ 10$^{-11}$ yr$^{-1}$), despite the fact that most of these galaxies are known to be star-forming systems.  We attribute their location in the figure to being entirely due to the fact that their H$\alpha$ SFRs (instantaneous SFRs) significantly underestimate the time-averaged recent SFRs as measured by their FUV fluxes.

In order to provide a more useful look at the sSFR$_{H\alpha}$ values for the SHIELD and comparison samples, we decided to prune the data set in order to retain only galaxies with SFR$_{H\alpha}$ values that were in reasonable agreement with their FUV SFRs.  Our methodology is best visualized by considering Figure~\ref{fig:ratiovsstellarmass}.  In this figure the H$\alpha$ SFRs for all galaxies above log(M$_{stellar}$) $\approx$ 8.5 can be considered to be fairly representative of their time-averaged recent SFRs.  The galaxies in this mass range are nearly all confined to values of log(SFR$_{H\alpha}$/SFR$_{FUV}$) between roughly $-$0.6 and 0.1.  Most of these galaxies are from the 11HUGS sample.  We adopt the lower limit of this range -- log(SFR$_{H\alpha}$/SFR$_{FUV}$) = $-$0.6 -- as an indicator of the point where SFR$_{H\alpha}$ become less representative of the time-averaged recent SFR.  We therefore impose a cut to our data (SHIELD as well as comparison galaxies) where we remove any galaxy with log(SFR$_{H\alpha}$/SFR$_{FUV}$) $<$ $-$0.6 and log(M$_{stellar}$) $<$ 8.5.

We remake our stellar mass vs. sSFR$_{H\alpha}$ plot using our excised data set in the right panel of Figure~\ref{fig:StellarMassVsssfr2}.  With the galaxies with low values of log(SFR$_{H\alpha}$/SFR$_{FUV}$) removed, the region occupied by the low-mass galaxies appears to be much more ordered.  All of the SHIELD galaxies with H$\alpha$ non-detections (open blue diamonds) have been removed, along with a number of H$\alpha$-detected SHIELD galaxies, FIGGS, 11HUGS, and LITTLE THINGS systems.  This cleaned version of the figure looks quite similar to the corresponding stellar mass vs. sSFR$_{FUV}$ plot from \citet[][(Figure 13)]{gormanous26}.  In particular, neither the sSFR$_{FUV}$ figure nor the revised sSFR$_{H\alpha}$ plot shown here have galaxies from either the SHIELD or the comparison samples with sSFR values $<$ 10$^{-11}$ yr$^{-1}$.

It is worth stressing that the cut imposed on the data, which is based on the SFR$_{H\alpha}$/SFR$_{FUV}$ ratio, does not simply remove all of the low-mass galaxies.  Indeed, some of the very lowest-mass galaxies present in the left panel of Figure~\ref{fig:StellarMassVsssfr2} are still present in the right panel.  This is consistent with the point made above in Section~\ref{sec:sfrratio} that departures from SFR$_{H\alpha}$ $\approx$ SFR$_{FUV}$ are not simply dependent on the mass of the galaxy, but rather are correlated with the SFR.  Several galaxies in the SHIELD and comparison samples with low stellar masses have SFR$_{H\alpha}$ measurements that are in good agreement with their SFR$_{FUV}$ values.

Our culled version of the stellar mass vs. sSFR$_{H\alpha}$ plot in Figure~\ref{fig:StellarMassVsssfr2} shows that nearly all of the galaxies in our combined SHIELD and comparison samples possess sSFR$_{H\alpha}$ values that fall between 10$^{-9}$ and 10$^{-11}$ yr$^{-1}$, while the galaxies have stellar masses that span five orders of magnitude.   For the most part, the SHIELD galaxies overlap the sSFR$_{H\alpha}$ values of the 11HUGS and FIGGS galaxies, while a substantial subset of the LITTLE THINGS galaxies stand out as having higher sSFR$_{H\alpha}$ values.  
Only a handful of galaxies exhibit higher star-formation efficiencies ($>$ 10$^{-9}$ yr$^{-1}$).   These include the SHIELD galaxy AGC 198691 (a.k.a. the Leoncino dwarf), which is the blue diamond located at log(sSFR$_{H\alpha}$) $\sim$ $-$9.0 yr$^{-1}$ and log(M$_{stellar}$) $\sim$ 5.9 M$_\odot$.  The paucity of galaxies with higher sSFR$_{H\alpha}$ is noteworthy, and was pointed out by \citet{gormanous26}.  Since all of the galaxies plotted in Figure~\ref{fig:StellarMassVsssfr2} are very local (most have distances under 13 Mpc), the lack of any extreme star-forming systems is no surprise given the low volume densities of galaxies with global starbursts in both dwarf galaxies \citep{jclee2009a} and Green Pea like galaxies \citep{brunker2020}.

\begin{figure*}
\centering
\includegraphics[width=6.0in]{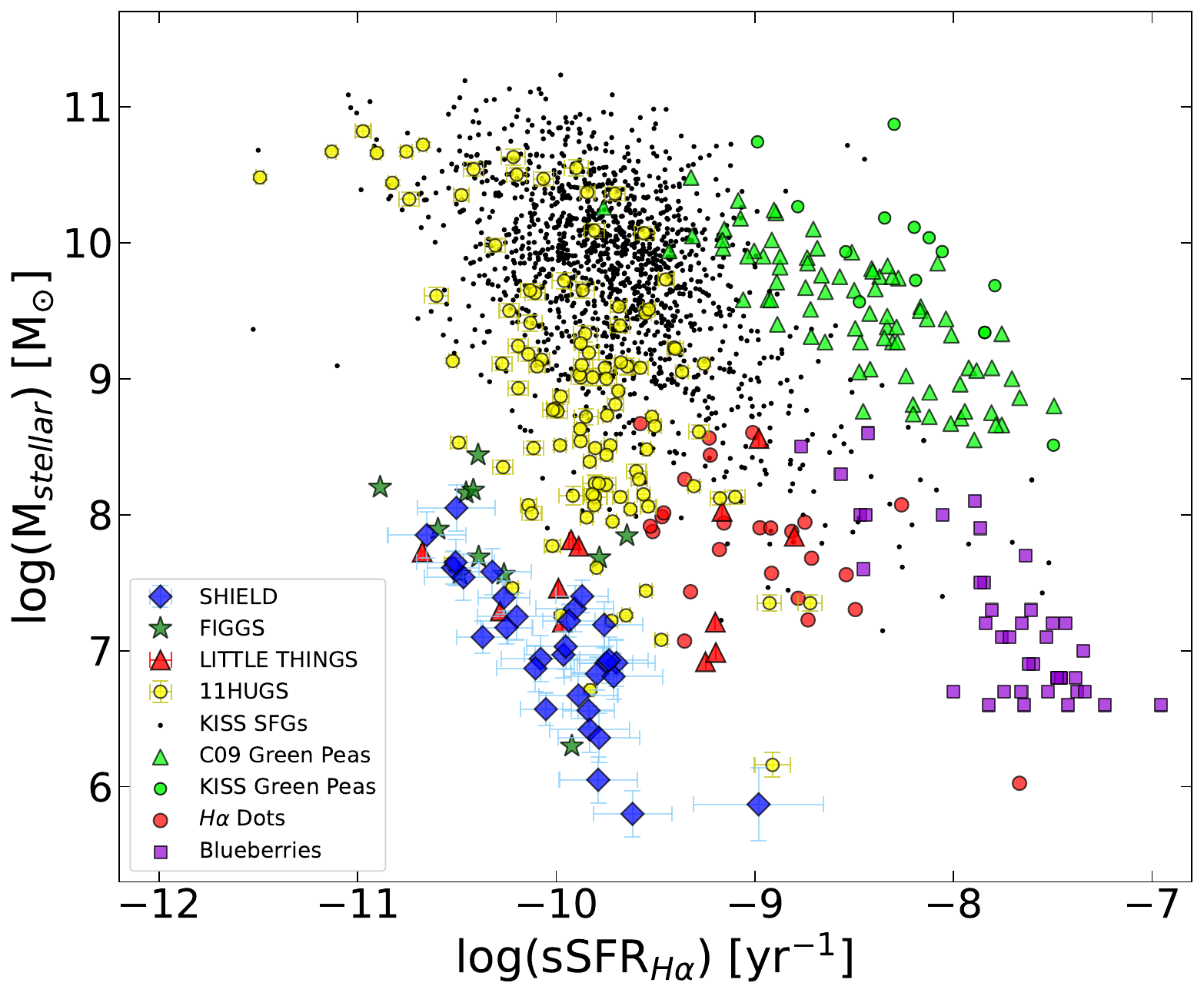}
\caption{Log-log plot of Stellar Mass vs.\ H$\alpha$ sSFR.  Here we combine the SHIELD galaxies and comparison samples used in the current paper as plotted in Figure~\ref{fig:StellarMassVsssfr2} with several samples of more extreme star-forming galaxies presented in \citet{hirschauer2022}, including Green Pea and Blueberry galaxies.  Details are provided in the text.  The SHIELD dwarfs and comparison sample galaxies from Figure~\ref{fig:StellarMassVsssfr2} are seen to have H$\alpha$ sSFR values that are between one and three orders of magnitude smaller than the extreme star-forming galaxies like Green Peas, BCDs, and bluberries.}
\label{fig:StellarMassVsssfr1}
\end{figure*}

In an effort to put the star formation properties of the SHIELD galaxies into better perspective, we present an expanded version of the stellar mass vs.\ sSFR$_{H\alpha}$ plot in Figure~\ref{fig:StellarMassVsssfr1}.   In this plot we present precisely the same galaxies shown in the righthand plot in Figure~\ref{fig:StellarMassVsssfr2}.
We retain the same symbol type and colors that have been used in all previous plots. To these galaxies we add systems with higher (in some cases extreme) sSFRs.  Small black dots represent generic H$\alpha$-detected star-forming galaxies (SFGs) from the KISS emission-line galaxy (ELG) survey \citep{salzer2000, salzer2001, gronwall2004, jangren2005}.  Green triangles and circles are Green Pea galaxies from \citet{cardamone2009} and \citet{brunker2020}, respectively.  At lower masses, the red circles are BCD galaxies detected in the H$\alpha$ Dot narrowband survey \citep{kellar2012, salzer2020, watkins2021, hirschauer2022}, while the purple squares are Blueberry galaxies from \citet{yang2017}.  We note that the total mass range of the newly added star-forming systems matches well with the galaxies used in the current study.

\begin{figure}
\centering
\includegraphics[width=3.35in]{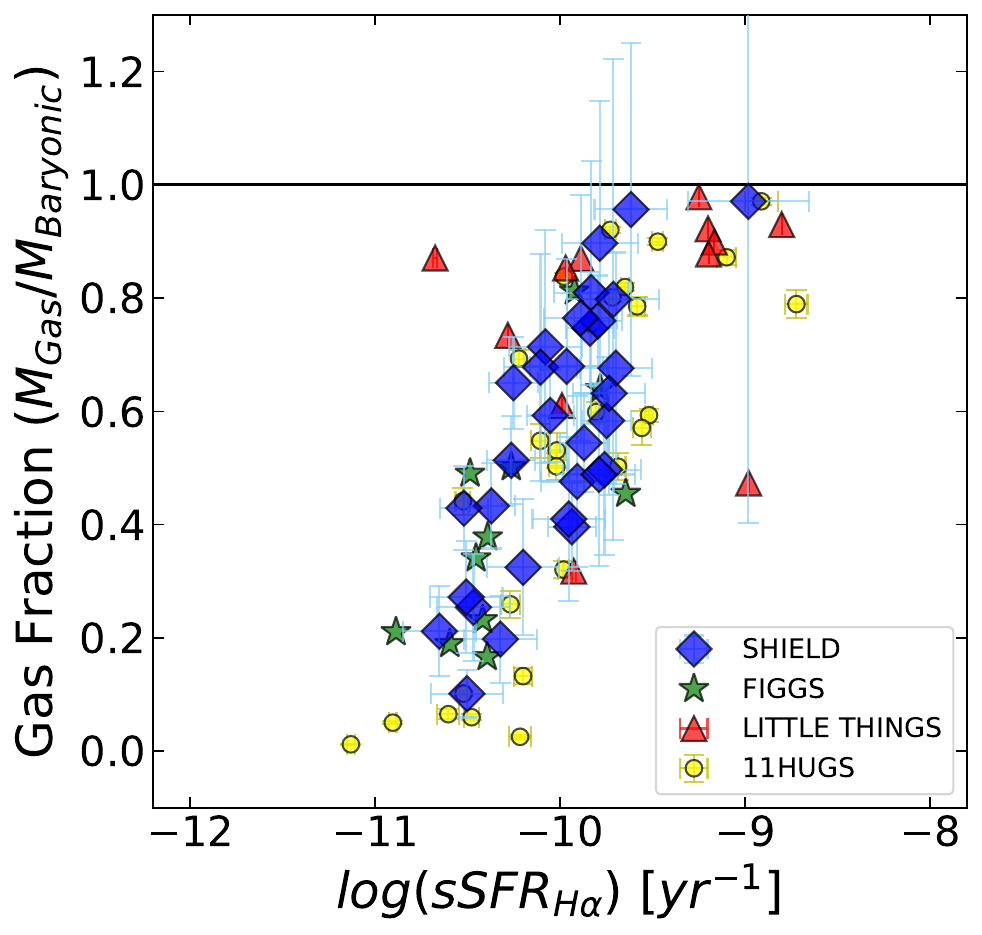}
\caption{Graph of baryonic gas mass fraction $(M_{gas} / M_{baryonic})$ vs.\ the logarithm of H$\alpha$ sSFR. The solid line indicates equality $(M_{gas} = M_{baryonic})$, which would be the gas for a starless galaxy.  We restrict the sample of galaxies plotted to those with SFR$_{H\alpha}$ values that are in reasonable agreement with their time-averaged recent SFRs (see Figure~\ref{fig:StellarMassVsssfr2}).  In contrast to Figure~\ref{fig:BaryonicFractionVssfr}, we see a significant correlation between the gas mass fraction and H$\alpha$ sSFR. }
\label{fig:BaryonicFractionVsssfr}
\end{figure}

As one might expect, there are clear offsets in sSFR$_{H\alpha}$ between the various samples.  The samples used for the current study (SHIELD, FIGGS, LITTLE THINGS, and 11HUGS) are all located in the local universe (within $\sim$15 Mpc).  As stressed above, while  they are all gas rich and have all had some level of star formation in the past few hundred Myrs, none of them would be considered to be an extreme starbursting system.  For the galaxies in the upper half of Figure~\ref{fig:StellarMassVsssfr1} (log(M$_{stellar}$) between $\sim$8.5 and 11) the 11HUGS galaxies are seen to overlap with the KISS SFGs, although they are offset, on average, by $\sim$0.5 dex to lower sSFRs.  The two samples of Green Pea galaxies are offset even more: the 11HUGS galaxies barely overlap with the Green Peas, and are offset on average by between 1.5 and 2.0 dex.   Green Pea galaxies with the highest sSFR values are a factor of 1000 times more active than the 11HUGS galaxies with the same mass but low sSFRs.

The dwarf galaxies shown in Figure~\ref{fig:StellarMassVsssfr1} (log(M$_{stellar}$) between $\sim$6.0 and 8.5) are similarly segregated into groups based on their sample selection.  The local dwarfs from SHIELD, FIGGS, and LITTLE THINGS are mainly located in the region with log(sSFR$_{H\alpha}$) $<$ $-$9.5, although there are a few 11HUGS and LITTLE THINGS dwarfs that extend to somewhat higher sSFR values.  The latter overlap with the BCDs from the H$\alpha$ Dot survey, which are all selected via H$\alpha$ emission in narrowband images.  Offset to even higher sSFR values are the Blueberry galaxies which, like the Green Peas, are selected using broadband photometry.

The SHIELD galaxies are seen to mostly inhabit a restricted region in Figure~\ref{fig:StellarMassVsssfr1}.  With the exception of AGC 198691, they all have log(sSFR$_{H\alpha}$) values between $-$9.6 and $-$10.7.   In the mass range 10$^{6.5}$ to 10$^{8.0}$ M$_\odot$ they are offset to lower sSFR$_{H\alpha}$ by a factor of $\sim$30 (1.5 dex) on average compared to the H$\alpha$ Dot BCDs and by a factor of $\sim$300 (2.5 dex) on average compared to the Blueberries.  Despite the fact that the SHIELD galaxies are strongly gas mass dominated, they are rather unremarkable in terms of their star formation efficiencies.  

Finally, we compare the baryonic gas mass fraction (M$_{gas}$/M$_{baryonic}$) to the logarithm of sSFR$_{H\alpha}$ in Figure~\ref{fig:BaryonicFractionVsssfr}.  This plot allows us to explore the degree to which the availability of gas correlates with star formation efficiency.  Again, we only plot galaxies that have instantaneous SFR values (SFR$_{H\alpha}$) that are in good agreement with their time-averaged SFRs (SFR$_{FUV}$).

Unlike many of our previous plots, we see that the galaxies from the various samples overlap significantly in Figure~\ref{fig:BaryonicFractionVsssfr}.  This is distinct from Figure~\ref{fig:BaryonicFractionVssfr}, which has the same y-axis but in that case the x-axis is not mass normalized.  In Figure~\ref{fig:BaryonicFractionVssfr}, the three dwarf galaxy samples plotted exhibit only modest overlap.   Similarly, in Figure~\ref{fig:StellarMassVsssfr2}, the x-axis is the same used here, but the y-axis is not mass normalized, leading to significant separation between the various samples.  Here, with both axes being mass normalized, the galaxies from the various samples overlap and suggest a significant correlation.

Figure~\ref{fig:BaryonicFractionVsssfr} shows a clear trend of increasing sSFR$_{H\alpha}$ with increasing gas mass fraction.  Although there are a few outliers, particularly among the LITTLE THINGS galaxies, the majority of the galaxies fall within a fairly tight relation. A similar trend was found in \citet{gormanous26} when sSFR$_{FUV}$ was used instead of sSFR$_{H\alpha}$ (see their Figure 14).  The tight relation seen here is in sharp contrast to Figure~\ref{fig:BaryonicFractionVssfr}, where there is absolutely no correlation between the baryonic gas mass fraction and SFR$_{H\alpha}$.

To summarize the results of this section, our analysis of the specific SFRs reveals that the SHIELD galaxies have rather ``normal" values of sSFR$_{H\alpha}$, despite being gas-dominated systems.  They overlap completely the galaxies in the comparison samples, which, like the SHIELD galaxies, are very local (distances mostly $\le$ 13 Mpc).  This point is further accentuated when we compare with more extreme star-forming galaxies (Figure~\ref{fig:StellarMassVsssfr1}) -- Green Peas, BCDs, and blueberries -- where these ``burstier" systems are offset from the SHIELD galaxies by between 1.5 dex (BCDs) and three dex (blueberries).  The SHIELD galaxies show a modest trend toward higher sSFR$_{H\alpha}$ with higher gas mass fraction (Figure~\ref{fig:BaryonicFractionVsssfr}), consistent with the simple-minded picture that galaxies with higher gas fractions are more efficient at making stars.  However, even galaxies with M$_{gas}$/M$_{baryonic}$ $>$ 90\% are not found to have high values of sSFR.  Very high gas mass fractions alone are not sufficient to cause galaxies to fall within the groups of extreme star-forming systems.


\section{Summary and Conclusions}\label{sec:summary}
We have presented new optical broadband and H$\alpha$ narrowband imaging, flux measurements, luminosities, and star-formation rates for the complete sample of 82 galaxies in the Survey of \ion{H}{1} in Extremely Low-mass Dwarfs (SHIELD). In addition to providing the first uniform census of {\it current} star formation activity across the full SHIELD sample, this work presents the first complete set of optical and continuum-subtracted H$\alpha$ images for all 82 SHIELD galaxies. H$\alpha$ emission is detected in 55 galaxies (67.1\% of the sample), while upper limits are derived for the remaining 27 systems. These observations complement the FUV-based SFR analysis presented in \citet{gormanous26}. Our principal results are summarized as follows:

\begin{itemize}

\item The SHIELD galaxies exhibit a wide range of optical morphologies and surface brightnesses despite being selected solely on the basis of their \ion{H}{1} content. The H$\alpha$ morphologies are generally simple, with the majority of H$\alpha$-detected galaxies containing only one or two identifiable H\,{\sc ii} regions. This result emphasizes the highly localized nature of current star formation in extremely low-mass dwarf galaxies.

\item Comparisons between H$\alpha$ and FUV SFRs confirm that the two tracers diverge systematically in the lowest-SFR systems. The departure from the relation SFR$_{\mathrm{H}\alpha} \approx$ SFR$_{\mathrm{FUV}}$ begins near $\log(\mathrm{SFR}_{H\alpha}) \approx -2.5$, below which galaxies exhibit progressively lower H$\alpha$-to-FUV SFR ratios. The SHIELD sample densely populates this low-SFR regime and clearly defines the strength of this departure.

\item The divergence between H$\alpha$ and FUV SFRs is also associated with galaxy mass. We show that the departure from equality (SFR$_{H\alpha}$ $\sim$ SFR$_{FUV}$) becomes apparent near $\log(M_{\rm stellar}) \approx 8.0$ M$_\odot$, and below this value galaxies are significantly more likely to exhibit strong deviations between their H$\alpha$ and FUV SFRs. However, the degree to which the instantaneous (H$\alpha$) SFR underestimates the time-averaged recent (FUV) value for galaxies below this mass limit is {\it not} correlated with stellar mass.  That is, at values of M$_{\rm stellar}$ $<$ 10$^8$ M$_\odot$, the observed ratio of SFR$_{H\alpha}$/SFR$_{FUV}$ can exhibit any value between 1.0 and 0.01, essentially independent of the stellar mass of the galaxy.
Several SHIELD galaxies with $\log(M_{\rm stellar}) \lesssim 7.0$ M$_\odot$ have H$\alpha$ SFRs that are in good agreement with their FUV SFRs.

\item Comparisons of H$\alpha$ SFR with stellar mass, \ion{H}{1} mass, baryonic mass, and absolute magnitude reveal approximately linear scaling relations above $\log(\mathrm{SFR}_{H\alpha}) \approx -2.5$. Of these quantities, the baryonic mass exhibits the tightest correlation with H$\alpha$ SFR and the smallest scatter. However, this predictive power is limited primarily to galaxies with $\log(M_{\rm baryonic}) \gtrsim 8.0$; below this threshold the relationship between galaxy mass and H$\alpha$ SFR breaks down.

\item The baryonic gas mass fraction, $M_{\rm gas}/M_{\rm baryonic}$, exhibits no correlation with H$\alpha$ SFR.  Galaxies at any given gas fraction span the complete range of H$\alpha$ SFRs exhibited by the SHIELD and comparison samples.  This result, which is consistent with the FUV-based analysis of \citet{gormanous26}, demonstrates that the availability of gas alone is not a reliable predictor of current star formation activity.  It is possible that the angular momentum content of the gas disk may help to control the conversion of gas into stars, allowing for a wide range of both gas mass fractions and SFRs \citep[e.g.,][]{obreschkow2016, lutz2018, mancerapina2021b}.

\item In contrast to the previous point, the baryonic gas mass fraction {\it does} exhibit a strong correlation with sSFR$_{H\alpha}$, in the sense that galaxies with higher $M_{\rm gas}/M_{\rm baryonic}$ have higher mass-normalized SFRs.  The efficiency or ``burstiness" of star formation does appear to scale with the gas fraction even though the total SFR does not.

\item The SHIELD galaxies tend to exhibit low H$\alpha$ specific SFRs.  Only a few of the galaxies in our analysis would be considered to have high sSFR$_{H\alpha}$ values, including the ultra-low metallicity system AGC 198691 (the Leoncino dwarf).  Comparison of the SHIELD galaxies with extreme star forming systems such as BCDs, Blueberries, and Green Peas emphasizes the rather unremarkable nature of the star formation activity within the SHIELD sample, despite the fact that they are all gas-rich systems.

\end{itemize}

Taken together with the results of \citet{gormanous26}, our findings support a picture in which star formation in extremely low-mass galaxies is highly stochastic. The SHIELD sample uniquely probes the low-mass parameter space where these effects become most pronounced, revealing a fundamental transition in galaxy behavior near $\log(M_{\mathrm{stellar}}) \approx 8.0$. Future analyses combining the H$\alpha$ imaging presented here with resolved \ion{H}{1} observations and optical spectroscopy will further constrain the interplay between gas content, chemical enrichment, and star formation in the lowest-mass gas-rich galaxies known.


\begin{acknowledgements}

D.G. would like to thank John and A-Lan Reynolds for their support of the Reynolds Post-Baccalaureate Fellowship, administered by the Indiana University Astronomy Department, which supported him during the completion of this project.  J.J.S. would like to express his thanks and gratitude to the College of Arts and Sciences at Indiana University for their ongoing support of research carried out in the Astronomy Department.  This project utilized archival images obtained by the GALEX satellite and made available to the community through the Mikulski Archive for Space Telescopes (MAST). We gratefully acknowledge the many scientists who worked to create and operate GALEX as well as the members of the MAST team who continue to make these valuable data available.  Finally, we wish to thank the anonymous referee for their many helpful suggestions.

\end{acknowledgements}

\vspace{5mm}
\facilities{WIYN, GALEX, MAST}




%
%


\bibliography{SHIELD_HA_SFR.bib}{}
\bibliographystyle{aasjournalv7}



\end{document}